\documentclass[technologies,article,accept,pdftex,moreauthors]{Definitions/mdpi} 
\firstpage{1} 
\pubvolume{13}
\issuenum{12}
\articlenumber{588}
\pubyear{2025}
\copyrightyear{2025}
\externaleditor{Hongying Meng and Daniele Giansanti} 
\datereceived{20 October 2025} 
\daterevised{24 November 2025} 
\dateaccepted{9 December 2025} 
\datepublished{13 December 2025} 
\hreflink{https://\linebreak \textls[-35]{doi.org/10.3390/technologies13120588}} 

\usepackage{amsmath,amssymb,amsfonts}
\usepackage{algorithmic}
\usepackage{textcomp}
\usepackage{multirow}
\usepackage{threeparttable}
\usepackage{bm}
\usepackage{array}
\usepackage{siunitx}
\usepackage{booktabs}
\usepackage{subcaption}
\usepackage{acronym}

\newacro{IMU}[IMU]{inertial measurement unit}
\newacroplural{IMU}[IMUs]{inertial measurement units}
\newacro{FSR}[FSR]{force sensing resistor}
\newacroplural{FSR}[FSRs]{force sensing resistors}
\newacro{SFO}[SFO]{smart foot orthosis}
\newacroplural{SFO}[SFOs]{smart foot orthoses}
\newacro{AFO}[AFO]{ankle-foot orthosis}
\newacroplural{AFO}[AFOs]{ankle-foot orthoses}
\newacro{IoT}[IoT]{internet of things}
\newacro{DFS}[DFS]{diabetic foot syndrome}
\newacro{PPG}[PPG]{photoplethysmography}
\newacro{EMG}[EMG]{electromyography}
\newacro{ESP}[ESP]{Espressif Systems Platform} 
\newacro{PCB}[PCB]{printed circuit board}
\newacro{BLE}[BLE]{Bluetooth Low Energy}
\newacro{LED}[LED]{light emitting diode}
\newacroplural{LED}[LEDs]{light emitting diodes}
\newacro{SMD}[SMD]{surface-mounted device}
\newacro{GUI}[GUI]{graphical user interface}
\newacro{THT}[THT]{through-hole technology}
\newacro{ERC}[ERC]{electrical rule check}
\newacro{UART}[UART]{universal asynchronous receiver transmitter}
\newacro{MQTT}[MQTT]{message queuing telemetry transport}
\newacro{ECAD}[ECAD]{Electronic Computer-Aided Design}
\newacro{CAD}[CAD]{computer-aided design}
\newacro{PLA}[PLA]{polylactid}
\newacro{TPU}[TPU]{thermoplastic polyurethane}
\newacro{SDK}[SDK]{software development kit}
\newacro{API}[API]{application programming interface}
\newacro{APK}[APK]{android package kit}
\newacro{AI}[AI]{artificial intelligence}
\newacro{GPS}[GPS]{Global Positioning System}
\newacro{UI}[UI]{user interface}
\newacro{mHealth}[mHealth]{mobile Health}
\newacro{LRA}[LRA]{linear resonant actuator}
\newacro{ERM}[ERM]{eccentric rotating mass motor}
\newacro{IC}[IC]{integrated circuit}
\newacro{DRC}[DRC]{design rule checks}
\newacro{COP}[COP]{center of pressure}
\newacro{MoCap}[MoCap]{motion capture system}
\newacroplural{MoCap}[MoCaps]{motion capture systems}
\newacro{I²C}[I²C]{inter-integrated circuit communication}
\newacro{IDE}[IDE]{integrated development environment}
\newacro{SUS}[SUS]{System Usability Scale}
\newacro{HCI}[HCI]{human-computer interaction}
\newacro{MCU}[MCU]{microcontroller unit}
\newacro{RQ}[RQ]{research question}
\newacroplural{RQ}[RQs]{research questions}
\newacro{SRQ}[SRQ]{supporting research question}
\newacroplural{SRQ}[SRQs]{supporting research questions}
\newacro{MRQ}[MRQ]{main research question}
\newacro{RM}[RM]{repeated-measures}
\newacro{ANOVA}[ANOVA]{Analysis of Variance}
\newacro{ART}[ART]{Aligned Rank Transform}

\Title{Smart Device Development for Gait Monitoring: Multimodal Feedback in an Interactive Foot Orthosis, Walking Aid, and Mobile Application}

\TitleCitation{Smart Device Development for Gait Monitoring: Multimodal Feedback in an Interactive Foot Orthosis, Walking Aid, and Mobile Application}

\Author{Stefan Resch 
 $^{1,2,}$*\orcidA{}, André Kousha 
 $^{1}$\orcidB{}, Anna Carroll $^{3}$\orcidC{}, Noah Severinghaus $^{4}$, Felix Rehberg $^{1}$, Marco Zatschker $^{1}$, \linebreak  Yunus Söyleyici $^{1}$\orcidD{} and Daniel Sanchez-Morillo $^{5,6}$\orcidE{}}

\AuthorNames{Stefan Resch, André Kousha, Anna Carroll, Noah Severinghaus, Felix Rehberg, Marco Zatschker, Yunus Söyleyici and Daniel Sanchez-Morillo}

\isAPAStyle{%
       \AuthorCitation{Resch, S., Kousha, A., Carroll, A., Severinghaus, N., Rehberg, F., Zatschker, M., Söyleyici, Y., \& Sanchez-Morillo, D.}
         }{%
        \isChicagoStyle{%
        \AuthorCitation{Resch, Stefan, André Kousha, Anna Carroll, Noah Severinghaus, Felix Rehberg, Marco Zatschker, Yunus Söyleyici, and Daniel Sanchez-Morillo.}
        }{
        \AuthorCitation{Resch, S.; Kousha, A.; Carroll, A.; Severinghaus, N.; Rehberg, F.; Zatschker, M.; Söyleyici, Y.; Sanchez-Morillo, D.}
        }
}

\address{%
$^{1}$ \quad Faculty of Computer Science and Engineering, Frankfurt University of Applied Sciences, \linebreak  60318 Frankfurt am Main, Germany
\\
$^{2}$ \quad Doctorate Program in Manufacturing, Materials and Environmental Engineering, Doctorate School (EDUCA), University of Cadiz, 11003 Cádiz, Spain \\
$^{3}$ \quad Department of Biomedical Engineering, University of Illinois Chicago, Chicago, IL 60607, USA
 \\
$^{4}$ \quad School for Engineering of Matter, Transport, and Energy, Arizona State University, Tempe, AZ 85287, USA
\\
$^{5}$ \quad Department of Automation Engineering, Electronics and Computer Architecture and Networks, \linebreak  School of Engineering, University of Cadiz, 11519 Cádiz, Spain
\\
$^{6}$ \quad Instituto de Investigación e Innovación Biomédica de Cádiz (INiBICA), 11009 Cádiz, Spain}

\corres{Correspondence: stefan.resch@alum.uca.es}

\abstract{
Smart assistive technologies such as sensor-based footwear and walking aids offer promising opportunities {for} gait rehabilitation through real-time feedback and patient-centered monitoring.
{While biofeedback applications show great potential, current research rarely explores integrated closed-loop systems with device- and modality-specific feedback.}
In this work, we present {a modular sensor-based system combining a smart foot orthosis and an instrumented forearm crutch to deliver real-time vibrotactile biofeedback.}
The system integrates plantar pressure and motion sensing, vibrotactile feedback, and wireless communication via a smartphone application.
We conducted a user study with eight participants to validate {the system’s feasibility for mobile gait detection and app usability, and to evaluate different vibrotactile feedback types across the orthosis and forearm crutch.}
{The results indicate that pattern-based vibrotactile feedback was rated as more useful and suitable for regular use than simple vibration alerts.
Moreover, participants reported clear perceptual differences between feedback delivered via the orthosis and the forearm crutch, indicating device-dependent feedback perception.}
{The findings highlight the relevance of feedback strategy design beyond hardware implementation and inform the development of user-centered haptic biofeedback systems.}
}


\begin{document}

\section{Introduction}
\label{sec:introduction}
Gait rehabilitation plays an important role in the recovery process after acute foot injuries and in the management of chronic foot-related conditions.
A variety of assistive devices are used to support these patients in their daily mobility, including orthopedic devices such as foot orthoses, insoles, bandages, and pressure-relieving footwear~\cite{ReviewOrthopedicFootwear}. 
In addition, mobility aids such as crutches, walking sticks, or canes are commonly used for temporary or long-term walking impairments~\cite{ReviewWalkingAids}.
While these devices are essential in stabilizing gait and reducing mechanical stress~\cite{ReducedPressure, Benefits-OrthopedicFootwear}, they are typically standardized and require customization to achieve optimal recovery outcomes~\cite{LackofPersonalization}, as they can measurably affect gait~\cite{gaitstudy}.
These limitations underscore the need for more individualized and adaptive orthopedic solutions that combine health monitoring with interactive feedback. 

In recent years, the integration of wearable electronics has contributed to various developments such as smart insoles, socks, and shoes equipped with sensor technologies for gait analysis and health monitoring~\cite{Salis.2023, Smart-Insoles-Review-2022, SmartFootwear.2024}. 
In contrast to conventional stationary systems, these mobile solutions enable gait analysis outside clinical settings and open up new possibilities for continuous monitoring of spatial, temporal, and spatio-temporal parameters~\cite{Treadmil-GoldStandard}.
These systems frequently rely on \acp{FSR} and \acp{IMU} to assess parameters such as plantar pressure distribution and gait characteristics (e.g., step cadence and foot strike patterns)~\cite{FSRandIMU}. 
Such technologies show promise in areas such as fall risk assessment, biofeedback interventions, and the prevention of foot ulcers~\cite{Foot-Temperature-Review.2022}. 
In particular, monitoring plantar pressure is crucial for patients with \ac{DFS}, where elevated sole pressure can impair wound healing and increase the risk of ulcer formation~\cite{Diabeticfoot}.
They also hold potential for injury prevention and early pathology detection~\cite{Insole-Systems-Overview-2022}.

{However,} orthopedic devices such as \acp{AFO}, which are used to stabilize gait and support recovery, {still rarely combine embedded sensing and feedback within an interactive system.
Integrating feedback modalities (such as haptic alerts triggered by pressure overload) could enhance the benefits of \ac{SFO} development, particularly for individuals with foot conditions~\cite{Resch-WalkingAid}.}
For example, diabetic neuropathy in the lower extremities impairs the sensation of pain and limits the detection of excessive plantar pressure~\cite{DiabeticNeuropathy}{, which can lead to} unnoticed injuries and increase the risk of serious deformities such as Charcot foot~\cite{CharcotFoot}.
Providing patients with intrinsic feedback could therefore positively impact gait recovery.
{While several conceptual prototypes have integrated electronics into orthoses~\cite{Smart-AFO-2023, Smart-Boot-2023, SmartAFO-Potential-EMG}, most solutions remain technically focused, without addressing how haptic feedback should be designed or delivered. 
There is a lack of empirical evidence on how feedback modality in smart footwear and walking aids influences user perception, perceived effectiveness, and intended use.
This is critical, as unclear or intrusive feedback can reduce acceptance even when technically accurate.}

To address the identified research gap, {we developed a modular sensor system to investigate the influence of haptic feedback modality and device-dependent feedback on user perception in smart footwear and walking aids.}
The system comprises three key components: (1) an \ac{SFO} equipped with \ac{IMU} and \ac{FSR} sensors as well as a vibration feedback module, (2) a sensor-integrated walking aid that communicates with the orthosis and includes its own \ac{IMU} and haptic actuators, and (3) a mobile smartphone application for health monitoring and feedback interaction. 
The system architecture enables real-time communication across all components and {provides the basis for evaluating closed-loop vibrotactile biofeedback strategies.}
{To ensure a valid basis for user perception testing, we examined the system’s feasibility for gait detection and the usability of the mobile application from a user interaction perspective. These validation objectives guided the following \acp{SRQ}:}

\begin{itemize}
    \item {SRQ1 
 (System Feasibility): How accurately does the developed \ac{SFO} measure plantar pressure and gait parameters compared to a clinical reference system, as a foundation for reliable real-time multimodal feedback?}
    \item {SRQ2 (Application Usability): How do users evaluate the usability of the accompanying mobile application for interpreting gait and pressure information within the integrated system?}
\end{itemize}

 {{Based}  on this, the core objective of this work is to investigate feedback strategy design for user-centered biofeedback, addressing the \ac{MRQ}:}
\begin{itemize}
    \item {MRQ (Feedback Evaluation): How do different vibrotactile feedback types (continuous vs.\ pattern-based), delivered through the orthosis and the walking aid, influence user perception in terms of noticeability, intrusiveness, perceived usefulness, and intended use?}
\end{itemize}

{This work contributes to the field of assistive technology by introducing a novel closed-loop, sensor-driven haptic feedback mechanism in a sensorized foot orthosis and instrumented forearm crutch, enabling real-time feedback that responds to gait dynamics and pressure distribution.
Our key research findings provide empirical insights that users significantly prefer pattern-based over simple vibrotactile feedback notifications and that device location influences user perception. 
These results offer concrete recommendations for designing effective, user-centered haptic rehabilitation systems and can serve as a basis for future patient-centered evaluations of multimodal feedback in \ac{HCI}.}

\section{Related Work}
\label{sec:relatedwork}
This section reviews related research in the areas of smart footwear for mobile gait analysis and health monitoring (Section~\ref{sec:RW1}), foot augmentation with multimodal feedback for patient–device interaction (Section~\ref{sec:RW2}), and sensor-integrated walking aids for mobility support (Section~\ref{sec:RW3}).
We conclude by summarizing current research gaps and outlining our contributions to the field.

\subsection{Smart Footwear for Mobile Gait Analysis and Monitoring Health Parameters} 
\label{sec:RW1}
The research field of sensor technology demonstrates that different types of sensors are used for measuring continuous physiological data depending on the application~\cite{Insole-Systems-Overview-2022}.
Based on the device type, sensor technologies can be integrated into various systems such as socks/textiles~\cite{Socks.2017, Socks-2019, Socks-2020, Socks-Shoes-Review-2020, Socks2022}, shoes~\cite{In-Shoe.2001, Shoes-Review.2017, Shoe-GaitPatterns.2020, Shoe-Review-2020, Shoes-IMU}, insoles~\cite{Insole.2016, Smart-Insoles-Review-2022, SmartInsoles.2012, SmartFootwear.2024, InsolesAutism}, or orthoses~\cite{Smart-AFO-2023, Smart-Boot-2023, IoT-Orthosis.2023, IoT-Orthosis.2024, IntelligentAFO-PressureSensing}.
In particular, smart insoles are developed for a broad range of applications, ranging from ambulatory gait assessment and home-based monitoring~\cite{InsolePCB, Insoles-HomeTrainingGaitAnalysis}, to gait rehabilitation after stroke~\cite{Insoles-Stroke}, fall risk detection~\cite{Insoles-FallDetection}, and activity classification using machine learning techniques~\cite{InsolesReview-ML, Insole-LowCost}.

A systematic literature review by Subramaniam et al.~\cite{Insole-Systems-Overview-2022} identified two primary health-related functions of smart insoles: (1) the measurement of plantar pressure distribution and (2) the analysis of gait and activity patterns.
Plantar pressure data are typically acquired using integrated \acp{FSR} for the assessment of foot stability, balance, body weight distribution, and early detection of pressure overload conditions.
A review by Razak~et~al.~\cite{15FSR-Sensors} recommended the use of 15 \ac{FSR} sensors to ensure sufficient coverage of plantar pressure distribution according to the individual foot anatomy. 
This recommendation is based on an anatomical segmentation of the foot sole into 15 regions: heel (areas 1–3), midfoot (4–5), metatarsals (6–10), and toes (11–15), which support most of the body load~\cite{15FSR-Original}.
However, the individual shoe size and foot anatomy need to be considered for proper sensor placement to prevent measurement errors~\cite{Placement-Error-FSR}.

Furthermore, the use of \ac{IMU} sensors (a combination of accelerometers and gyroscopes) enables the measurement of gait characteristics such as step length, cadence, walking speed, and swing phase duration~\cite{GaitAnalysisIMU}. 
These systems are used for performance tracking of athletes~\cite{Insole-Systems-Overview-2022} or to enable mobile gait analysis~\cite{Insole-GaitAnalysis}.
A single IMU sensor embedded in the insole can provide sufficient data for this purpose~\cite{MotivationIMU}.
However, the literature reports inconsistencies regarding the optimal sensor placement for precise data extraction, as signal drift and measurement errors may occur depending on the location of the sensor~\cite{IMUplacement}.
As a result, calibration remains a critical challenge in smart insole development~\cite{Insole-Systems-Overview-2022}.
To minimize mechanical load and improve signal quality, placement in the midfoot region is generally recommended~\cite{Insole-Systems-Overview-2022}.

Beyond \ac{IMU} and \ac{FSR}-based gait analysis, additional sensing modalities have been explored to extend the functionality of wearable systems. 
These include \ac{GPS} tracking for spatial gait analysis~\cite{GPStracking}, foot temperature monitoring for early detection of \ac{DFS}~\cite{Foot-Temperature-Review.2022, DiabeticUlcer-Insoles}, and physiological measurements such as heart rate via \ac{PPG} sensors~\cite{PPG-signal-2018} and muscle activity using \ac{EMG} electrodes in textile-based wearables~\cite{Socks-2019}. 
However, \ac{PPG} and \ac{EMG}-based approaches remain largely limited to research prototypes and have not yet been adopted in commercial devices. 
Other studies have also investigated sensing technologies for obstacle detection, including ultrasonic sensors, ToF cameras, and radar-based systems, particularly in applications for visually impaired users~\cite{Review-Shoe-Obstacles}.
While these alternative sensing modalities offer promising extensions for specific use cases, the majority of current smart insole systems rely on the integration of \ac{IMU} and \ac{FSR} sensors.
Due to its efficiency and widespread adoption, the combination is frequently used in smart insole systems~\cite{FSR-IMU-Position}.
This configuration enables accurate gait detection~\cite{Review-GaitAnalysis, IMUandFSR-GaitDetection} and simultaneous acquisition of kinetic (pressure) and kinematic (motion) data~\cite{FSRandIMU, IMUandFSR-Limitations}.
Typically, data from \ac{IMU} and \ac{FSR} sensors are acquired using a portable microcontroller and transmitted wirelessly (e.g., via Bluetooth) to an external \ac{GUI}~\cite{InsolePCB, Wireless}.

To ensure the reliability of gait-related measurements derived from wearable systems, their performance must be validated against established reference standards.
A review by Hutabarat et al.~\cite{Treadmil-GoldStandard} emphasized that such wearable gait systems need to be validated against proven standard gait measurement systems to ensure data accuracy.
For this purpose, instrumented treadmills and \acp{MoCap} represent the gold standard for validating gait analysis systems~\cite{GoldStandardMoCap, Treadmil-GoldStandard}.
This approach has already been applied in several validation studies, for instance by comparing wearable systems to a stationary Zebris treadmill~\cite{Validation-Zebris-Treadmill} or GAITRite force platforms~\cite{eShoe-vs-Forceplate, GaitWear}.
Overall, recent research highlights the importance of comparative validation studies with established gait analysis systems to strengthen the credibility and reliability of findings in wearable gait assessment~\cite{Review-NeedforFutureResearch}.
In conclusion, despite considerable advances in this research field, key challenges remain regarding data accuracy, system validation, and participant-centered~evaluation.

\subsection{Foot Augmentation with Multimodal Feedback for Patient Device Interaction} 
\label{sec:RW2}
Smart wearable systems such as insoles, socks, or orthoses are often connected to external devices (e.g., smartphone applications, smartwatches, or other interfaces) to enable continuous mobile health monitoring and interactive feedback~\cite{Insole-Systems-Overview-2022}.
A systematic review by Saboor et al.~\cite{SmartphonePotential} highlighted the growing potential of smartphone applications for mobile gait analysis, particularly in remote health contexts, while also noting ongoing challenges related to accessible and user-friendly \ac{UI} design.
In this context, Resch~et~al.~\cite{Resch2023} developed a \ac{mHealth} application for smart insoles on the basis of focus groups with patients and experts.
In a follow-up study, functional mockups were evaluated in a usability test with 30 participants, which confirmed the user-friendliness and provided recommendations for future design improvements~\cite{Resch2024}.
The resulting design recommendations, integrated features, and overall app concept provide a valuable basis for the continued development of smart insole applications and the integration of feedback~modalities.

Beyond interface design, wearable systems increasingly incorporate real-time feedback mechanisms (through visual, acoustic, or haptic signals) typically via smartphone apps or connected devices to support user interaction.
For instance, vibration cues have been applied to indicate pressure overload~\cite{Vibration-Insoles}. 
It was shown that vibration feedback applied within the shoe can significantly influence gait behavior in healthy participants~\cite{Oates2017}, and vibrotactile biofeedback triggered by FSR input reduced overpronation duration by 50\%~\cite{Vibrotactile-Insole}.
Furthermore, auditory feedback has also been investigated, such as acoustic signals triggered by in-sock pressure sensors for gait correction~\cite{AppInsoles-Auditory}, or auditory stimulation via insoles for patients with Parkinson’s disease~\cite{AuditoryInsoles}.
These approaches are particularly relevant for patients with impaired peripheral sensation, such as individuals with \ac{DFS}, for whom smartwatch-based alerts and real-time gait correction via smart insoles have been proposed~\cite{Insole-HapticFeedback}.  
Several studies have applied multimodal feedback approaches and shown that they influence gait symmetry, posture and pressure distribution~\cite{App-Feedback, Oates2017, Vibrotactile-Insole, Elvitigala}.
In an experimental study, Redd et al.~\cite{App-Feedback} compared visual, acoustic, and vibrotactile feedback modalities in healthy participants and found that both visual and haptic cues significantly improved gait symmetry.
Despite the demonstrated potential of smartphone applications and feedback modalities, further validation is needed regarding user perception and system~effectiveness.

\subsection{Smart Walking Aids and Applied Sensor Technologies}
\label{sec:RW3}
Walking aids play a crucial role in daily mobility support for patients with walking impairments.
Recent research has explored how these devices can be enhanced through sensor technologies to support health monitoring, user interaction, and feedback-based interventions~\cite{Review-WalkingAID, technologies13080346}. 
Various smart walking aid prototypes have been proposed, including instrumented sticks~\cite{SmartStick}, canes~\cite{SmartCane-Impaired}, and forearm crutches~\cite{Forearm-Crutch}.
Many of these systems have been developed specifically for users with visual impairments~\cite{Stick-visuallyImpaired, Stick-VisuallyImpaired2, Stick-VisImpaired3, AugmentedCane}.
These systems typically contain sensors such as \acp{IMU} for fall detection or for motion tracking~\cite{Cane-and-Insoles, AngularVelocity}, \acp{FSR} embedded in handles to detect gait events~\cite{InstrumentedCane-FSR}, GPS modules for positional tracking~\cite{Cane-VisuallyImpairedGPS}, and ultrasonic sensors for obstacle detection~\cite{Stick-visuallyImpaired, Stick-VisuallyImpaired2, Stick-VisImpaired3}.  

Beyond data acquisition, several studies have explored the integration of feedback mechanisms to enhance user support~\cite{AugmentedCane, Forearm-Crutch-Reha, Haptic-Feedback-Cane}.  
For example, Merrett et al.~\cite{Forearm-Crutch-Reha} demonstrated the potential of biofeedback in forearm crutches using acoustic signals based on FSR data to support optimal weight distribution.  
Schuster et al.~\cite{Haptic-Feedback-Cane} showed that haptic feedback delivered via the handle of a cane can reduce knee adduction moments in individuals with knee osteoarthritis.
In addition, a smart stick with integrated SOS functionality has been developed that connects to smartphone applications~\cite{SOS-Stick}.

While individual sensor-based walking aids have demonstrated promising results, integrated systems that combine smart walking aids with intelligent footwear for foot augmentation remain rare and are typically limited to specific application scenarios.  
For instance, Suman et al.~\cite{WalkingAidandShoes} proposed a system in which a smart cane communicates with smart footwear, using ultrasonic sensors in the shoe and auditory feedback via the cane handle.  
Another example is the use of force-sensitive crutches to assist lower-limb exoskeleton users by capturing ground force data to enable active support~\cite{Exoskeleton-Crutch}.
Furthermore, a robotic walking cane was developed that receives \ac{COP} data from four \acp{FSR} integrated in an insole to support fall detection~\cite{Cane-and-Insoles}.
However, direct communication between walking aids and orthopedic devices, such as sensor-based foot orthoses, has not yet been realized.  
In this context, Resch et al.~\cite{Resch-WalkingAid} proposed a conceptual system that enables a walking aid to communicate with a smart AFO through vibration feedback in the handle to indicate plantar pressure overload.
Nevertheless, the proposed system remains a concept without implementation or empirical evaluation.

Overall, while sensor-based walking aids and feedback modalities offer high potential, most systems remain in the prototype stage and lack systematic validation or integration into broader rehabilitation frameworks~\cite{technologies13080346}.  

\subsection{Research Gap}
\label{sec:RW4}
In summary, various systems for mobile foot health monitoring have been developed, primarily based on \acp{IMU} and \acp{FSR}, often in combination with wireless communication to external devices such as smartphone applications.  
While smart insoles have been extensively studied, their integration into orthopedic devices such as \acp{AFO} remains limited.
Similarly, sensor-based walking aids show promising results for gait support and biofeedback, but most remain standalone prototypes without integration into more comprehensive assistive systems.  
Although haptic feedback has proven to be beneficial for improving user performance, there is no tested approach to date that combines it with sensor-based footwear and walking aids in a unified system.

Related work lacks concepts and empirical evidence for the combined use of smart walking aids and foot monitoring technologies.
However, this integration offers future potential for improving patient monitoring, supporting real-time gait analysis and enabling multimodal feedback in healthcare and everyday use.
Building on these findings, our work addresses these gaps by developing and validating an integrated system that combines a smart foot orthosis, an instrumented walking aid, and a smartphone application for real-time gait monitoring and interactive feedback.
By addressing these gaps, we contribute to the advancement of assistive device technologies, mobile gait analysis, and user-centered feedback systems.

\section{Method}
This section describes the overall system architecture and the technical implementation of the smart orthosis, the instrumented walking aid, and the smartphone application. 
It is structured as follows: Section~\ref{3.1} provides a system overview, Section~\ref{3.2} presents the hardware architecture and concept development, and Section~\ref{3.3} outlines the software architecture and data processing for each subsystem.
Finally, Section~\ref{3.4} presents the experimental setup of the prototype validation study, including apparatus, procedure, and measurements with data analysis.

\subsection{System Overview}
\label{3.1}
\begin{figure}[H]
    \centering
    \begin{adjustwidth}{-\extralength}{0cm}
        \includegraphics[width=\linewidth]{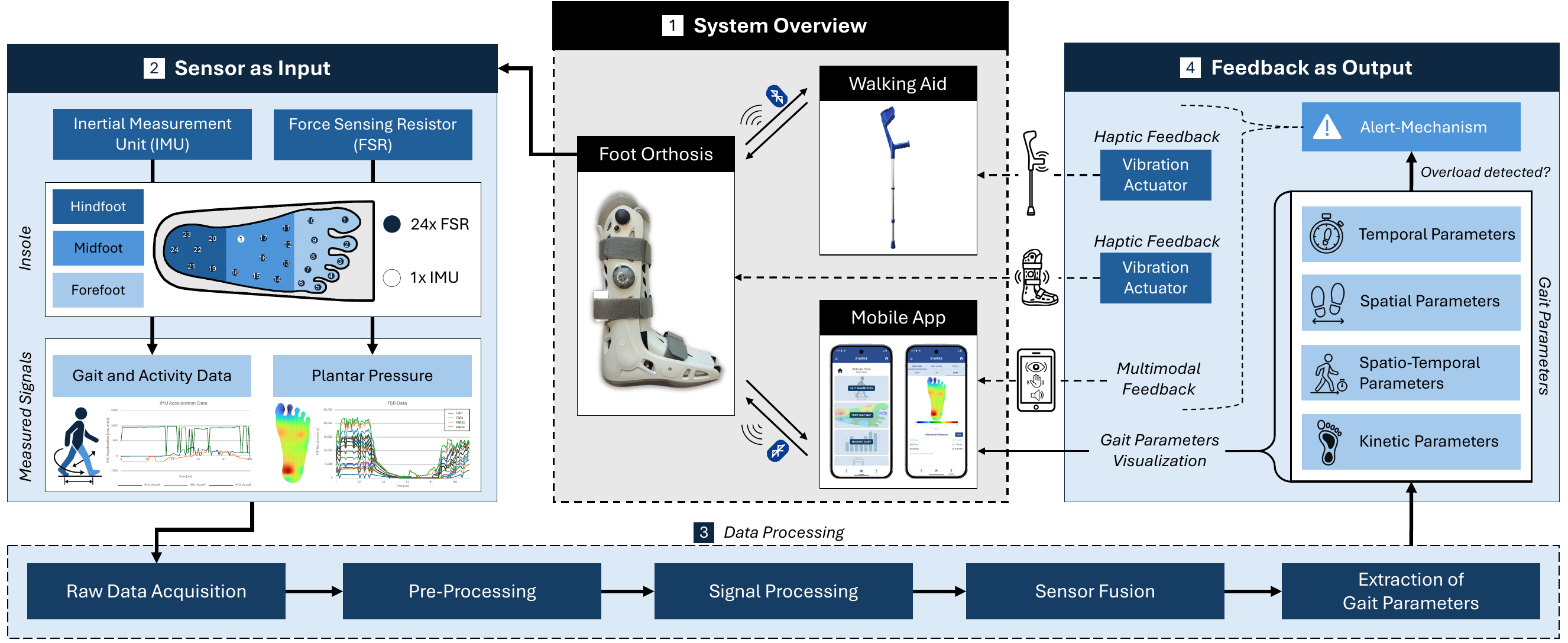}
    \end{adjustwidth}
    \caption{Overview of the sensor-based assistive system consisting of a smart orthosis with integrated FSR and IMU sensors, vibration feedback, and its communication with a walking aid and an mHealth application for interactive gait monitoring.}
    \label{SystemOverview}
\end{figure}

Figure~\ref{SystemOverview} provides an overview of the system architecture, showing the interaction between the \ac{SFO}, the instrumented walking aid, and the mobile application for data visualization and feedback control.
The SFO is based on a standardized lower-leg orthosis (AIRCAST\textsuperscript{\textregistered} AIRSELECT™ Elite Walker~\cite{Enovis} by ORMED GmbH, a company of Enovis, Freiburg, Germany), also known as a controlled ankle motion boot, which can be worn universally on either foot. 
A sensor concept consisting of multiple \acp{FSR} and a single \ac{IMU} is integrated into a 3D-printed insole to capture plantar pressure and motion data.  
The insole is designed as a modular component, with adjustable size and shape to accommodate different orthotic devices and foot sizes.

Sensor fusion and real-time data processing enable the extraction of gait parameters and the detection of critical loading patterns.  
To support active user guidance, haptic feedback is provided via a vibration motor embedded in the orthosis, which is triggered in response to excessive plantar pressure.  
A custom-designed \ac{PCB}, equipped with an on-board \ac{MCU}, enables wireless communication via \ac{BLE} with both the mobile application and the instrumented walking aid.
The forearm crutch is based on a standardized design (ISO 11334-1:2007)~\cite{NORM} and includes a custom-designed \ac{PCB} equipped with an \ac{MCU}, an \ac{IMU} sensor, two vibration modules, and a rechargeable battery.
All components are embedded in the handle to enable synchronized haptic feedback and motion sensing.
In addition, a corresponding \ac{mHealth} application developed in Android Studio provides a graphical interface for visualizing gait parameters and plantar pressure data. 

\subsection{Hardware Architecture and Concept Development}
\label{3.2}
\subsubsection{Smart Orthosis}
\begin{figure}[H]
\begin{adjustwidth}{-\extralength}{0cm}
\centering
\subfloat[\centering]{\includegraphics[width=0.75\linewidth]{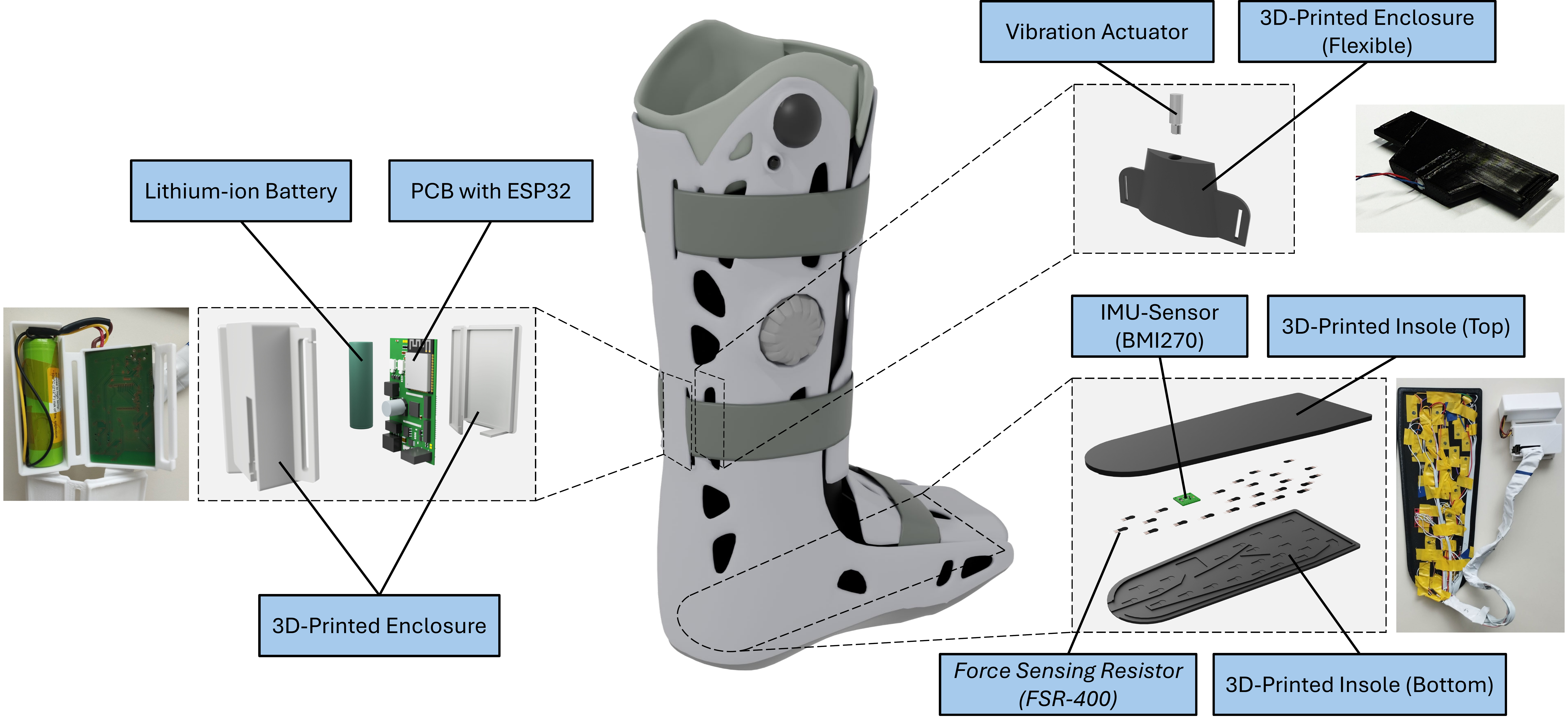}}
\hfill
\subfloat[\centering]{\includegraphics[width=0.23\linewidth]{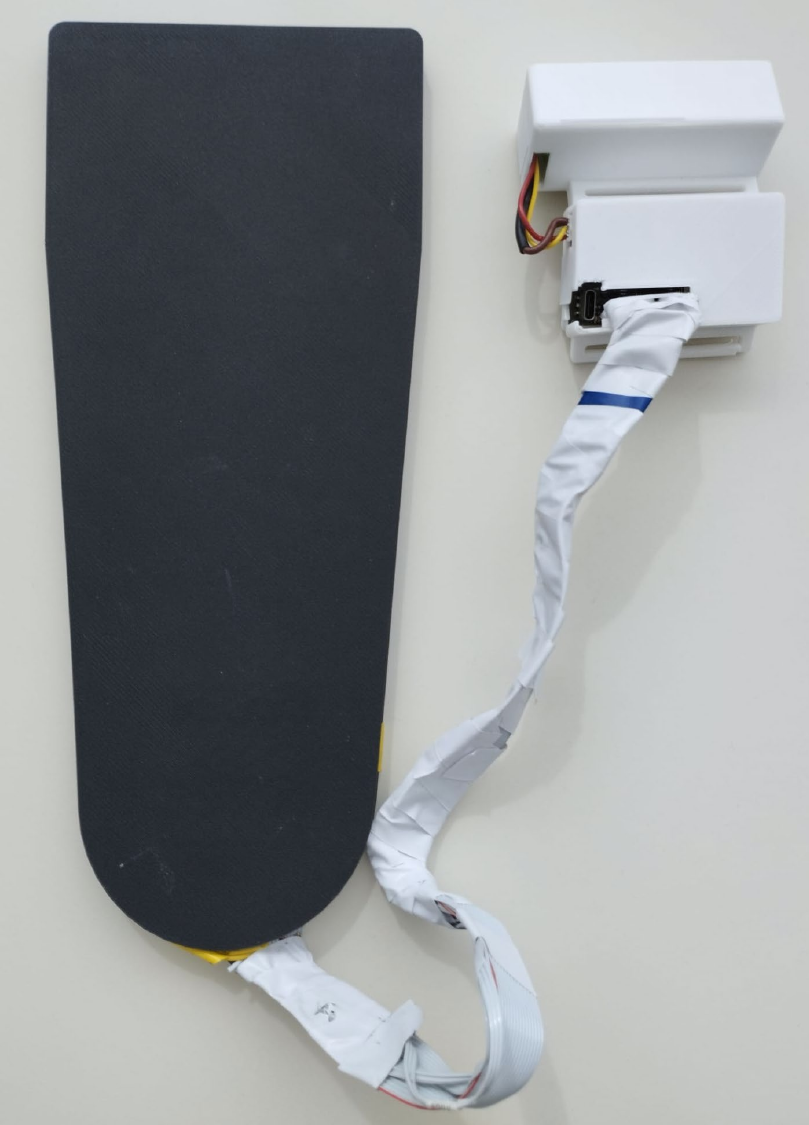}}\\
\end{adjustwidth}
\caption{(\textbf{a}) Overview of the orthosis prototype and assembled components; (\textbf{b}) Final assembled prototype of the smart insole.\label{figure2}}
\end{figure}

A schematic overview of the \ac{SFO} prototype is presented in Figure~\ref{figure2}a and the assembled insole with integrated sensor components is shown in Figure~\ref{figure2}b. 
The \ac{SFO} is based on a modular insole integrated into a standard lower-leg orthosis. 
An optional external vibration module can be placed at varying heights along the calf using the holes of the orthosis, allowing flexible positioning for haptic feedback.
The insole includes multiple \acp{FSR} and an \ac{IMU} sensor, connected via a flat cable to an external \ac{PCB}, which is enclosed in a 3D-printed housing mounted on the posterior strap of the orthosis.
The enclosure was designed using the \ac{CAD} software Autodesk Fusion 360 (v.2603.0.86) and fabricated from \ac{PLA} with a 0.4 mm nozzle (HT extruder) using the Flashforge Creator 4 3D printer.
The insole consists of a two-part structure, comprising a bottom and a top layer, both fabricated from flexible \ac{TPU} using the extruder F with 0.4 mm nozzle size.
The geometry replicates the original insole of the AIRCAST\textsuperscript{\textregistered} orthosis to ensure proper fit and alignment{, without affecting wearing comfort.}
The design enables adaptation for the contralateral healthy foot by modifying the insole shape and enclosure mounting to fit standard footwear. 
It also supports future integration into other orthotic systems tailored to individual foot anatomies.

The modular insole was designed to include the recommended minimum of 15~\acp{FSR}~\cite{15FSR-Sensors} (Model 400 short tail, Interlink Electronics~\cite{InterlinkElectronics}) to capture plantar pressure distribution.
The placement of the \acp{FSR} is based on the anatomical structure of the foot and covers key plantar regions, including the heel, forefoot, metatarsal arch, and toes.
The layout supports extensions up to 24 \acp{FSR}, which may be beneficial for applications involving foot deformities or more detailed regional pressure mapping.
Each sensor has an active area of 5.6 mm in diameter, a thickness of 0.3 mm, and a force sensitivity range of 0–20 N, offering fast response times and high reliability for dynamic pressure measurements~\cite{interlink_fsr_guide}.

\vspace{-3pt}

For motion tracking, a 6-axis \ac{IMU} (BMI270, Bosch~\cite{Bosch}) was integrated into the insole.
The \ac{IMU} was positioned near the metatarsal arch to minimize pressure exposure.
This sensor was selected based on key technical requirements for wearable gait analysis systems and is commonly used in fitness trackers and smart wearables, such as smart shoes for step counting and activity recognition~\cite{bosch2023bmi270}.
The ultra-low power sensor combines a 3-axis accelerometer (sampling frequencies from 25 Hz to 6.4 kHz) and a 3-axis gyroscope (0.78~Hz to 1.6 kHz) within compact dimensions of 0.8 mm $\times$ 3 mm $\times$ 2.5 mm.

A custom \ac{PCB} (1.55 mm thickness, 35 \si{\micro\meter} copper) with a total size of 60 $\times$ 35 mm was designed using Altium Designer (v.25.1.2) as \ac{ECAD} software.
A four-layer design was chosen to achieve a compact and reliable layout. 
The top layer was used for signals and components, the inner layers for ground planes, and the bottom layer for additional signal routing, allowing clear separation of signal paths, power supply, and grounding.
The \ac{SFO} is powered by a 3.7 V, 2600 mAh lithium-ion battery (Emmerich ICR-18650NQ-SP) integrated with a fuse, a shutdown switch, and a battery management system (MCP73838).
A buck-boost converter regulates the supply voltage to a stable 3.3~V.
To enable Wi-Fi, Bluetooth, and \ac{BLE} communication, a 2.4 GHz ESP32-WROOM-32E-N8 \ac{MCU} with an integrated \ac{PCB} antenna from Espressif Systems was used, which is widely utilized in \ac{IoT} applications.
Additional components include a USB-C charging port and a multicolor \ac{LED}, which indicates the connection state, battery level, and charging status.
Electrostatic discharge protection for electronic components was implemented in accordance with IEC 61000-4-2 and IEC 61340-5, using protection diodes on critical inputs such as the USB interface.
A 32-channel analog multiplexer was implemented to enable the integration of up to 24 \acp{FSR}, while keeping signal paths short and maintaining robust ground routing to preserve signal~integrity.

An optional interface for vibration feedback was implemented using a transistor-based low-side switch controlled via pulse-width modulation.
This simple and efficient solution was chosen over more complex driver ICs or \ac{I²C}-based vibration modules to reduce system complexity and footprint.
This approach allows for flexible use of an \ac{ERM}.
To generate a sufficiently strong tactile signal, an \ac{ERM} vibration actuator (Model TC-13324656, TRU Components) operating at 2.4 V and 250 mA was selected. 

To minimize space, components were placed and routed according to system requirements and manufacturer guidelines.
The final design was verified using \ac{ERC} and \ac{DRC}.
Board assembly was performed manually using solder paste and a heating plate for \ac{SMD} soldering, followed by manual mounting of \ac{THT} components.
The assembled board was visually inspected and functionally tested to ensure proper soldering and reliable system performance.
Additionally, detailed battery tests confirmed the functionality of the integrated battery management system, including discharge and charging profiles. 
The estimated runtime without vibration feedback was approximately 35.5 h, with a full recharge time of 6--7 h depending on the initial state of discharge.


\subsubsection{Smart Walking Aid}
\textls[-15]{To enable active haptic feedback through an assistive device, a forearm crutch was selected as the basis for sensor integration, as such walking aids are commonly used in combination with \acp{AFO} and offer a practical interface for feedback delivery.
We developed a modular integration concept that can be transferred to various walking aids, including sticks, canes, or walkers, provided the handle design allows for modification without compromising ergonomics.
This is essential, as major modifications to the handgrip require careful consideration of weight distribution and ergonomic design to avoid negative impact.
For example, sensor placement along the shaft could alter the center of gravity, which may require new validation/certification of the walking aid~\cite{Cane-Modifications}.
Therefore, our design specifically targeted integration without major modifications to the handgrip.
In this implementation, we used a standardized aluminum forearm crutch with a plastic grip and an integrated reflector.
After removing the reflector, the inner cavity of the handle provided a usable space of 25 mm in diameter and 115 mm in length for the integration of the electronic components.}

A custom-designed four-layer \ac{PCB} (1.55 mm thickness, 35 \si{\micro\meter} copper) with a total size of 90 $\times$ 21 mm was developed to ensure compact placement and signal integrity.
The board integrates a BMI270 \ac{IMU} for motion sensing, an ESP32-WROOM-32E \ac{MCU} (2.4 GHz Wi-Fi, Bluetooth) for wireless communication via \ac{BLE} with the \ac{SFO}, and a Type-C USB connector for programming and battery charging.
Power is supplied by an IFR 14500J AA lithium iron phosphate LiFePO4 battery (600 mAh, 3.2 V) from Keeppower, managed by a Microchip MCP73123 battery charge controller. 
A charge status \ac{LED} was connected to a status pin for visual indication.
To enable vibrotactile feedback, two DRV2605L haptic driver \acp{IC} from Texas Instruments were integrated, allowing the control of two \ac{ERM} or \ac{LRA} vibration modules. 

\begin{figure}[H]
\begin{adjustwidth}{-\extralength}{0cm}
\centering
\subfloat[\centering]{\includegraphics[width=0.47\linewidth]{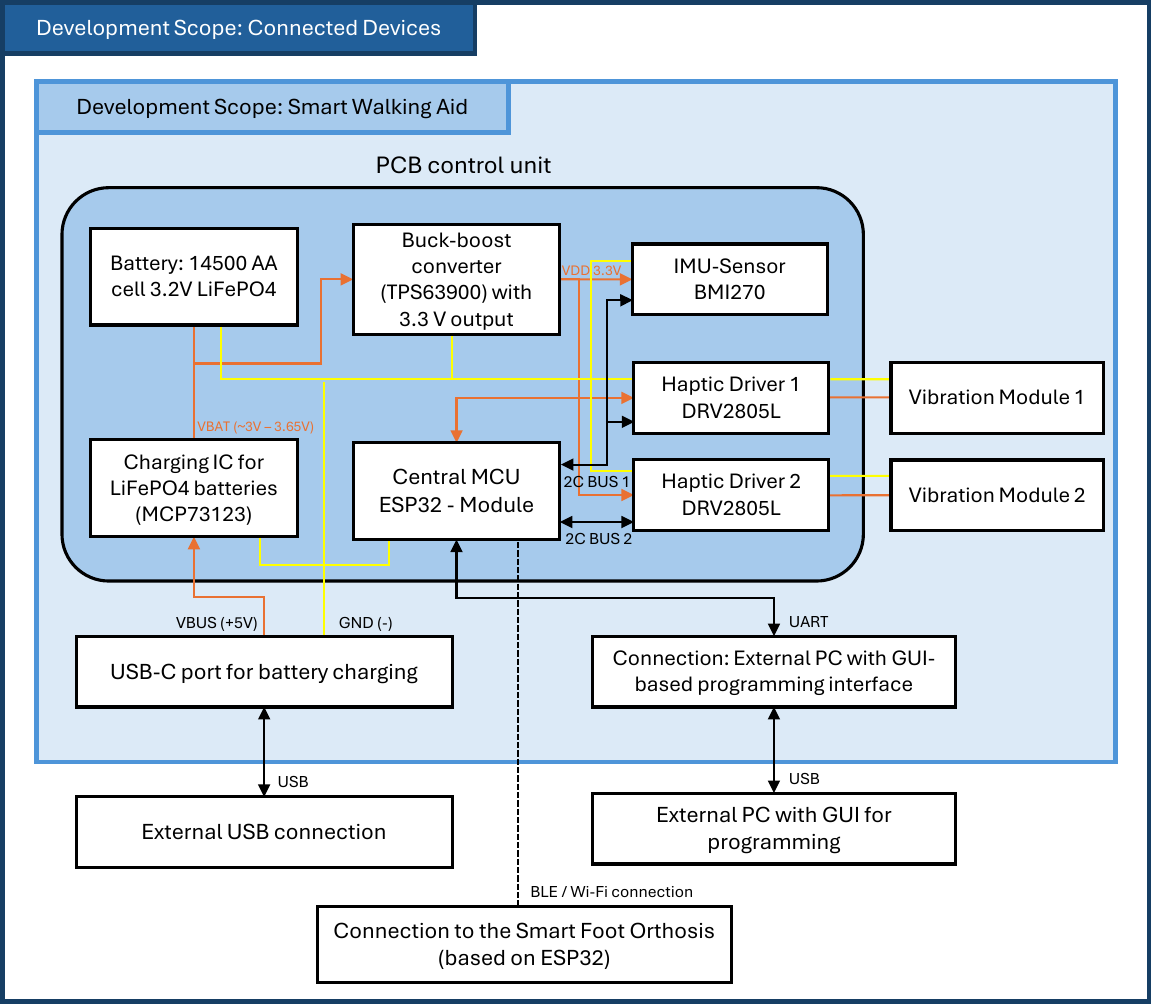}}
\hfill
\subfloat[\centering]{\includegraphics[width=0.51\linewidth]{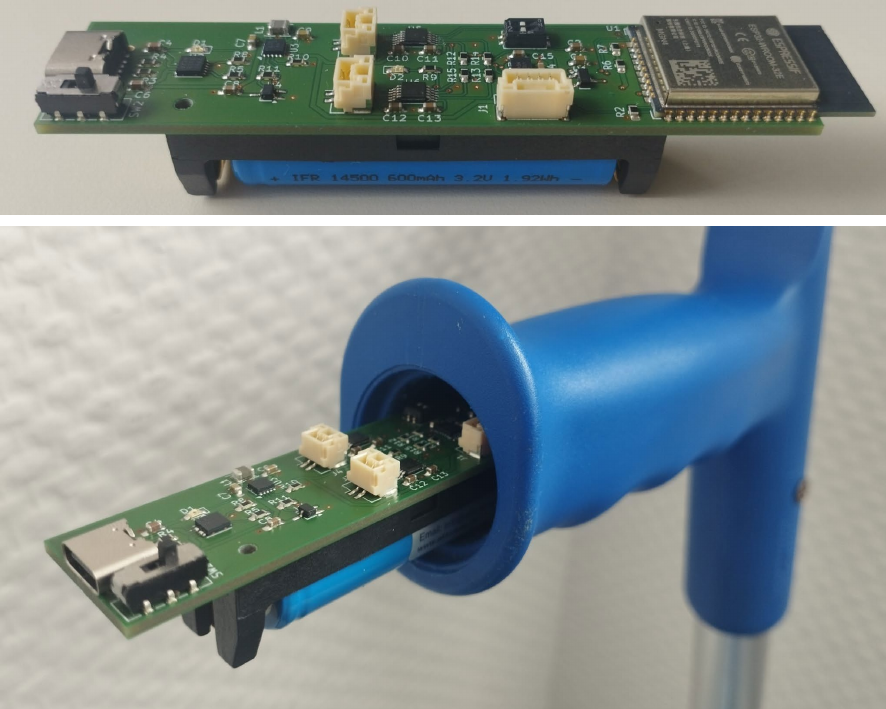}}\\
\end{adjustwidth}
\caption{(\textbf{a}) Hardware block diagram of the smart walking aid; (\textbf{b}) Final assembled prototype with integrated electronics.\label{figure3}}
\end{figure}

The USB connector, \ac{MCU}, and battery holder were symmetrically arranged along the central axis of the \ac{PCB} to ensure balanced layout and optimal use of space.
The \ac{MCU} and USB connector were placed on the front side, while the battery was positioned on the back side in accordance with the manufacturer’s mounting guidelines. 
A \ac{UART} interface was included to allow communication with an external PC and \ac{GUI} during development.
Before fabrication, \ac{ERC} and \ac{DRC} were performed to verify the circuit design and layout.  
The \ac{PCB} was then assembled using a manual pick-and-place process, followed by reflow soldering in a three-zone convection oven (Mistral 260) for \ac{SMD} components.  
Additional \ac{THT} components were soldered manually in a subsequent step.  
After assembly, the boards underwent functional testing and visual inspection to ensure proper soldering and system performance.
To ensure optimal positioning and mechanical stability within the handle, the assembled \ac{PCB} was mounted inside a custom 3D-printed enclosure made of rigid \ac{PLA}. 
To enable charging without disassembly, an access was added to the back of the crutch, providing external access to the USB-C port.
An overview of the hardware architecture is provided in Figure~\ref{figure3}a, and the assembled prototype is shown in Figure~\ref{figure3}b.
%
%


\vspace{-9pt}

\subsection{Software Architecture and Data Processing}
\label{3.3}
\subsubsection{Smart Orthosis---Mobile Gait Analysis}
\begin{figure}[H]
    \centering
    \begin{adjustwidth}{-\extralength}{0cm}
        \includegraphics[width=\linewidth]{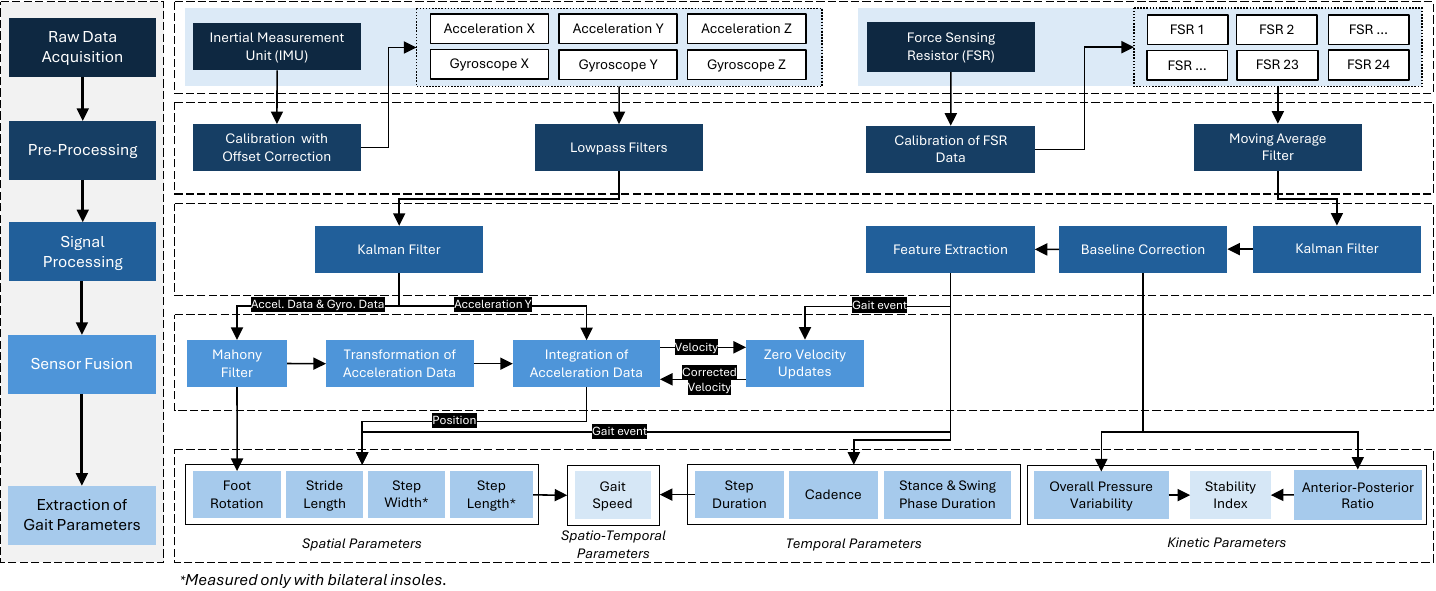}
    \end{adjustwidth}
    \caption{{Overview}  of the data processing for gait data extraction of the smart foot orthosis.}
    \label{SensorFusion}
\end{figure}

Figure~\ref{SensorFusion} presents the \ac{SFO} data processing architecture, structured into five stages: (1)~raw data acquisition, (2) pre-processing, (3) signal processing, (4) sensor fusion, and (5)~extraction of gait parameters.  
Each step is described in detail{; see Appendix~\ref{appendix:A}.}

\subsubsection{Smart Walking Aid---Haptic Feedback and Motion Data}
\textls[-15]{The ESP32 \ac{MCU} receives processed gait event data from the \ac{SFO} via wireless communication using either \ac{BLE} or Wi-Fi.
The firmware was developed in C/C++ using the Arduino \ac{IDE} and enables the control of two actuators to generate vibration patterns depending on pre-defined overloads.
In the current prototype implementation, relative pressure thresholds were defined as fixed percentage values of the user’s body weight to ensure consistent triggering conditions during system testing, similar to Mariani et al.~\cite{mariani2013quantitative}, who used a threshold of 5\% of body weight for gait phase detection.
This constant threshold design served to demonstrate the feedback functionality under controlled conditions and was not intended to represent clinically calibrated loads.
In future implementations, these thresholds could be dynamically personalized and adapted per gait phase (stance vs. swing) based on baseline load measurements obtained during standing and normal walking.
However, such thresholds are individual to each patient and thresholds should be defined and validated by medical experts to ensure the detection of clinically relevant overload conditions~\cite{Thresholds}.
Vibration signals are emitted using a configurable pulse-based pattern.
Depending on the detected parameter both the intensity and interval of the vibration can be dynamically adapted.
This customization allows targeted feedback for specific gait abnormalities, such as in-toeing or out-toeing patterns, where intensity can be increased with foot rotation angles.
The integrated \ac{IMU}, initialized with the same configuration as in the \ac{SFO}, continuously records motion data for activity recognition.
This data is also streamed to the smartphone application.
To extend the functionality, these data could be used in future implementations for activity classification or fall detection.}

\subsubsection{Smartphone Application}

The design of the smartphone application builds upon the main functions proposed by Resch et al.~\cite{Resch2023} for a smart insole app dedicated to foot health monitoring.  
Building on these proposed functions, our application implements key features such as a heatmap visualization of plantar pressure based on the \ac{FSR} sensors, real-time gait characteristics, and a walking diary displaying step counts and activity-related health status ratings.  
Additional modules include a knowledge database with a search function for articles on foot health, a training area featuring video tutorials for foot-related exercises, and a map for locating local medical specialists.  
The app also enables monitoring of the communication status between the \ac{SFO} and the walking aid.

\begin{figure}[H]
    
    \begin{adjustwidth}{-\extralength}{0cm}
\centering
        \includegraphics[width=0.9\linewidth]{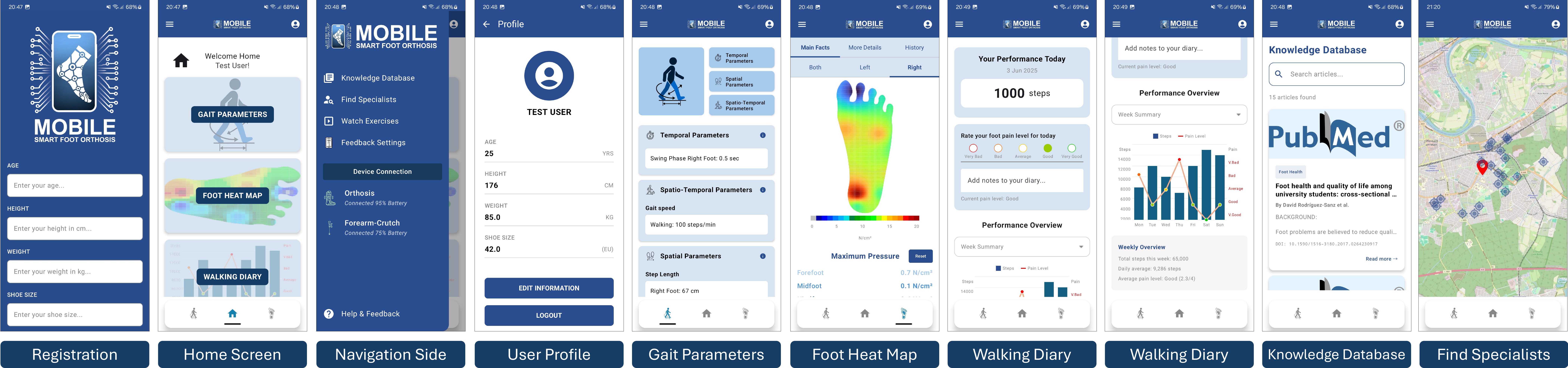}
    \end{adjustwidth}
    \caption{{Screenshots}  of the developed mHealth app showing key interface components, including account creation, main menu navigation, plantar pressure heatmap, gait diary, exercise tutorials, and access to foot health resources.}
    \label{MobileApp}
\end{figure}

For pressure data visualization, a custom heatmap fragment shader was implemented.
Its output was based on the number of \acp{FSR} used and colored using the Turbo colormap developed by Google~\cite{Google}.  
The application was developed in Android Studio~\cite{AndroidStudio} (version 2024.2.1) using Kotlin as the main programming language.
We utilized Jetpack Compose~\cite{JetpackCompose} and its integrated \ac{UI} libraries from Material2 and Material3, enabling flexible design of forms, buttons, and icons.  
Wireless communication was implemented via Wi-Fi, Bluetooth, and MQTT protocols to ensure real-time data transmission between devices.  
The prototype was deployed as an \ac{APK} on an Android 14 device using \ac{SDK} version 35.  
Figure~\ref{MobileApp} shows selected screenshots of the main functionalities of the developed \ac{mHealth} application. 

\subsection{Prototype Validation Study: Experimental Setup}
\label{3.4}
A technical feasibility and validation study was conducted with eight participants to evaluate the integrated system in terms of both objective measurement accuracy and subjective user experience.
The primary goal of this proof-of-concept investigation was to verify the system’s functional reliability and multimodal feedback performance under controlled laboratory conditions, rather than to focus on algorithmic novelty or benchmark testing of individual algorithms, or to conduct a clinical study.
{The study comprised two supporting validation objectives (system feasibility and application usability) and one primary objective (feedback evaluation).}
\begin{enumerate}
    \item {System Feasibility:} Validation of the \ac{SFO} plantar pressure and spatial/temporal gait data against an instrumented treadmill.
    \item {Application Usability:} User experience evaluation of the smartphone app based on the \ac{SUS}~\cite{SUS}.
    \item {Feedback Evaluation:} Assessment of two feedback mechanisms {applied within the \ac{SFO} and forearm crutch} using a customized questionnaire.
\end{enumerate}
In the following subsections, the study design, apparatus, procedure and tasks, as well as the applied measures and data analysis, are documented.

\subsubsection{Study Design}
\begin{figure} [H]
    \includegraphics[width=\linewidth]{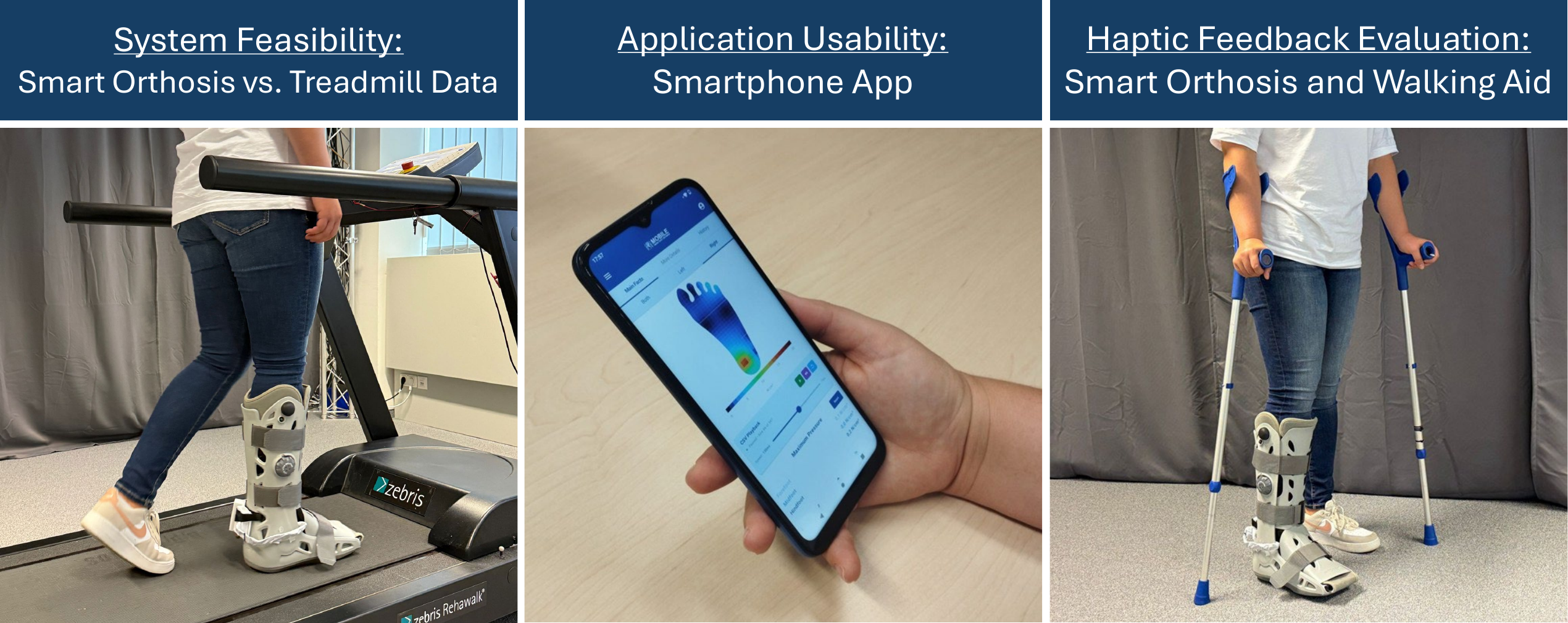}
    \caption{{Visual} 
 overview of the experimental validation process including orthosis testing, usability assessment, and haptic feedback evaluation.}
    \label{Task}
\end{figure}

The study was conducted in three parts, each with a specific validation objective.
An overview of the experimental validation and evaluation setup is shown in Figure~\ref{Task}.

\textls[-15]{To evaluate the measurement accuracy of the \ac{SFO}, an experimental validation was conducted against a clinical gait analysis reference system (Zebris FDM-T treadmill).
The independent within-subject variable \textit{Measurement System} included two levels: \textit{SFO} and \textit{Zebris Treadmill}.
Dependent variables included key spatial (step and cycle length), temporal (stance, swing, and cycle phase durations), and kinetic (plantar pressure distribution)~parameters.}

The smartphone app was tested to assess perceived usability using the standardized \ac{SUS} questionnaire. 

To evaluate the influence of haptic feedback on user perception, participants tested both the \ac{SFO} and the crutch, each equipped with one vibration module.
The study was based on a $2 \times 2$ factorial within-subject design with two independent variables: \textit{System} (two levels: \textit{Orthosis} and \textit{Crutch}) and \textit{Feedback Type} (two levels: \textit{Continuous} and \textit{Pulsed Pattern}).
Dependent variables were based on subjective ratings regarding noticeability, intrusiveness, regular usage, and perceived usefulness.
A $2 \times 2$ balanced Latin square design was applied to counterbalance the conditions and avoid sequence effects~\cite{BalanceLatinSquare}.

\subsubsection{Apparatus}
The prototype systems (\ac{SFO}, instrumented forearm crutch, and smartphone app) were used according to the previously described specifications and deployed in the study using MQTT for real-time data communication between the devices.
The 3D-printed insole was designed to fit EU shoe sizes 38–43, covering the medium shoe size range in both women and men.
To meet the minimum sensing requirements with optimal sensor placement, the insole of the orthosis was equipped with 15 \acp{FSR}.
Furthermore, a single \ac{ERM} 3.7V DC motor (model: JYCL0610R2540) from Jie Yi Electronics Limited was used as vibration actuator, integrated both into the orthosis and the handle of the crutch. 
This configuration was chosen to assess whether a single actuator can provide sufficient haptic feedback to be reliably perceived.
The app was executed on a Motorola Moto G8 Power Lite 64 GB (Android 10).

For the evaluation of motion and pressure data acquired from the \ac{IMU} and \ac{FSR} sensors, we used an instrumented treadmill, which is considered the gold standard for validating gait analysis systems.  
Specifically, a RehaWalk\textsuperscript{\textregistered} FDM-THPL-S-2i instrumented treadmill (Zebris Medical GmbH, Isny, Germany) was used to measure spatial and temporal gait parameters. 
The system is based on an h/p/cosmos pluto treadmill and integrates 3120 capacitive sensors ($\pm$0.05\% measurement accuracy), arranged in an 80 $\times$ 39 grid layout, covering a sensor area of 101.6 × 49.5 cm.
Data were recorded at a sampling rate of 100 Hz (similar to the \ac{SFO}), within a sensor range of 1--120 N/cm².
Zebris FDM software (version 2.0.4) was used to run gait analysis and export CSV raw data.
The software ran on a Windows 10 Pro system with an AMD Ryzen 5900X 12-core processor, 3.70 GHz, a GeForce RTX 3070 graphics card, and 32 GB RAM.

\subsubsection{Procedure and Tasks}
The study was conducted under controlled laboratory conditions without external influences.
First, participants were informed about the study procedures and that they could withdraw or discontinue participation at any time. 
After providing informed consent, they completed a survey about their demographic data, foot conditions, previous experience with assistive devices, and mHealth applications.
Furthermore, participants were asked to rate their familiarity with the use of smartphone healthcare apps.
Following this, the \ac{SFO} was provided to each participant as a fully preassembled system. 
An integrated vacuum-based mechanism was used to ensure optimal adaptation to the individual leg shape.
After a brief instruction, each participant completed five walking trials on the Zebris treadmill at a fixed speed of 1.5 km/h.
Each trial lasted 10 s to ensure the recording of at least one full gait cycle, typically capturing multiple cycles for averaging and comparison across multiple measurements.
Participants were instructed to walk as normally as possible without holding onto the treadmill handles.

After the walking trials, the walking aid prototype was introduced and tested both while standing and walking, in combination with the \ac{SFO}.
Haptic feedback was evaluated for both systems under predefined overload conditions to demonstrate feedback alerts and to ensure identical vibration triggering conditions across all participants.
Two feedback type modalities (continuous and pulsed patterns) were activated in randomized order for each participant.
Following this, participants completed a questionnaire consisting of quantitative Likert items to assess noticeability, intrusiveness, regular usage, and perceived usefulness.
Additionally, participants were asked to indicate their preferred vibration alert type in a single-choice question.

Finally, participants completed a smartphone usability task and assessed the app using the \ac{SUS} questionnaire.
An optional certificate of participation was offered after the study.

\subsubsection{Measurement and Data Analysis}
A quantitative measurement approach was chosen to evaluate the overall system and its individual subsystems.
Descriptive and inferential statistical analyses were conducted using R (version 4.2.2) with the rstatix package~\cite{R}.
The following paragraphs describe the corresponding measurements and data analysis in detail.

{\paragraph{System Feasibility: Accuracy of Plantar Pressure and Gait Parameters (Smart Orthosis)}}
Both systems recorded gait events and pressure-related data simultaneously. 
Manual triggering was performed during the swing phase of the right leg (foot orthosis) to synchronize both systems at the moment of initial right foot contact.
Each trial lasted for 10 s, recorded at a sampling rate of 100 Hz in both systems, resulting in five datasets per participant.
{The Zebris FDM software} 
 enabled the export of pressure data and gait characteristics were aggregated as mean and standard deviation values per trial.
The definitions of the respective gait parameters are provided in Table~\ref{tab:gait_parameters}.

To enable a valid comparison, the raw data from the \ac{SFO} were processed equally: gait characteristics were computed per trial and then aggregated on the participant level into mean and standard deviation values. 
Descriptive statistics were used to summarize the data, and inferential statistics were applied using paired t-tests to examine significant differences between both measurement systems.

\begin{table}[H]
\centering
\caption{Overview of analyzed gait parameters grouped by parameter type.}
\label{tab:gait_parameters}
\newcolumntype{C}{>{\centering\arraybackslash}X}
\begin{tabularx}{\textwidth}{LLL} 
\toprule
\textbf{Parameter Type} & \textbf{Parameter} & \textbf{Unit} \\
\midrule
\multirow{3}{*}{Temporal} 
    & Stance Duration & s \\
    & Swing Duration & s \\
    & Cycle Duration & s \\
\midrule
\multirow{1}{*}{Spatial} 
    & Stride/Cycle Length & m \\
\midrule
\multirow{1}{*}{Kinetic} 
    & Mean Plantar Pressure & N/cm$^2$ \\
\bottomrule
\end{tabularx}
\end{table}

For pressure data, the full time-series dataset was used for the \ac{SFO}, while the treadmill provided aggregated data for one step.
However, a direct quantitative comparison of force or pressure values between the treadmill and orthosis systems is not possible, due to fundamental differences in sensor configuration and measurement principles. 
The treadmill measures plantar pressure at the interface between the orthosis sole and the treadmill surface, while the orthosis records forces between the foot and the insole. 
As a result, the absolute values and spatial resolution of the pressure data differ between systems. 
Therefore, pressure data were visualized as heatmaps and force–time curves to qualitatively assess whether gait events could be reliably identified from the force signal~progression.

{\paragraph{Application Usability: Evaluation of Smartphone App Usability}}
Usability of the smartphone application was assessed using the standardized \ac{SUS}, which is frequently used to assess user experience of interactive systems such as smartphone apps.
The questionnaire consists of ten items rated on a five-point Likert scale ranging from 1 (\textit{strongly disagree}) to 5 (\textit{strongly agree}). 
Even-numbered items were reverse-scored and calculated according to the specifications by Brooke~\cite{SUS}.
The total score was calculated and converted into a value between 0 and 100 as a quantifiable measure of overall usability. 
Descriptive statistics were used to analyze the \ac{SUS} scores.\\


{\paragraph{Feedback Evaluation: User Perception of Haptic Feedback (Orthosis and Walking~Aid)}}

To assess the user experience and perceived effectiveness of the haptic feedback system, the participants rated four statements on a 7-point Likert scale~\cite{LikertScale}.
The items addressed noticeability, intrusiveness, regular usage, and perceived usefulness, with response options ranging from 1 (\textit{strongly disagree}) to 7 (\textit{strongly agree}).

The questionnaire items included:
\begin{itemize}
    \item Noticeability: ``\textit{The haptic feedback was clearly noticeable during use.}''
    \item Non-Intrusiveness ``\textit{I found the feedback comfortable and non-intrusive.}''
    \item Regular Usage ``\textit{I could imagine using such feedback during everyday mobility tasks.}''
    \item Usefulness ``\textit{The feedback would be useful in real-life situations (e.g., alerting me of incorrect foot placement or walking posture).}''
\end{itemize}

Descriptive statistics were applied to summarize the participants’ responses and identify trends in agreement levels.
A two-factorial \ac{RM} \ac{ANOVA} was used to assess the effect of system and feedback type on noticeability, intrusiveness, regular usage, and perceived usefulness.

{
\subsection{Participants}
The study was conducted in accordance with the data privacy regulations of our institution and received ethical approval from the German Society for Nursing Science (No. 23-027).
Eight participants were recruited through institutional mailing channels and provided informed consent before participation. 
To participate in the study, individuals were required to meet the following inclusion criteria:
\begin{itemize}
    \item Footwear size: European shoe size between 38 and 43.
    \item Current or past foot-related conditions, including either acute or chronic conditions
    \item Current or previous use of foot-related orthopedic products, such as shoe insoles, foot orthoses, bandages, or foot casts.
    \item Right foot affected or bilateral condition (but not solely left foot), as the sensor insole was configured specifically for the right foot.
\end{itemize}
Prior to study participation, all individuals completed a brief pre-screening to confirm that these criteria were met.
In total, 8 participants (4 female, 4 male) were recruited.
The age of the participants ranged between 24 and 72 years (\textit{M} = 39.75, \textit{SD} = 17.92).
Participants’ educational backgrounds included four with vocational training, three with a bachelor's degree, and one with a master's degree.
Occupations were distributed as follows: services and sales (\textit{n} = 2), engineering (\textit{n} = 2), other fields (\textit{n} = 2), clerical (\textit{n} = 1), and health-related professions (\textit{n} = 1).
The shoe size ranged from EU 38 to EU 43 (\textit{M} = 40.37, \textit{SD} = 1.85).
All participants reported having a current or past foot condition.
Four had previous conditions (such as a broken leg, sprained ankle, pes planus, or torn ligament).
Two participants reported chronic conditions (foot malalignment and arthrosis), one participant had an acute condition (sprained ankle), and one reported both an acute (sprained ankle) and chronic condition (torn ligament and previous leg fracture).
All participants indicated that they had previously used orthopedic devices, including insoles (\textit{n} = 7), ankle braces or bandages (\textit{n}~=~3), and foot orthoses (\textit{n} = 2).
Furthermore, four participants already used forearm crutches as medical walking aids.
In total, three participants had previous experience with a mobile health app.
Participants rated their familiarity with the use of smartphone healthcare apps on a scale from 1 (\textit{not at all familiar}) to 5 (\textit{extremely familiar}), with an average rating of 2.38~(\textit{SD} = 1.60).
All participants completed the study and were included in the~analysis.}

\section{Results}
\label{sec:Results}
{
The results section first presents the comparative accuracy validation between the \ac{SFO} and the instrumented treadmill (Section \ref{4.1}), followed by the usability assessment of the smartphone application (Section \ref{4.2}), and concludes with the evaluation of the haptic feedback of both the orthosis and the walking aid (Section \ref{4.3}).
}

\subsection{System Feasibility: Accuracy of Plantar Pressure and Gait Parameters (Smart Orthosis)}
\label{4.1}

\subsubsection{Gait Data Measurement}
To assess the reliability of temporal and spatial gait parameters, the accuracy of the \ac{SFO} measurements was evaluated against the treadmill system across all trials. 
The results showed that temporal parameters achieved mean accuracies above 85\%, except for swing phase duration, which reached only 60\%.
For the spatial parameter, the mean accuracy was 83.4\%, despite a high standard deviation.
The descriptive results are summarized in Table~\ref{tab:GaitAccuracy}.

\begin{table}[H]
\caption{Mean and standard deviation of gait parameter accuracy of the smart foot orthosis (SFO) compared to treadmill measurements across all participants.}
\label{tab:GaitAccuracy}
\newcolumntype{C}{>{\centering\arraybackslash}X}
\begin{tabularx}{\textwidth}{llcc} 
\toprule
\textbf{Parameter Type} & \textbf{Parameter} & \textbf{Mean Accuracy} & \textbf{SD Accuracy} \\
\midrule
\multirow{3}{*}{Temporal} 
    & Stance Phase Duration & 96.7\% & $\pm$3.1 \\
    & Swing Phase Duration  & 60.3\% & $\pm$12.6 \\
    & Cycle Phase Duration  & 85.2\% & $\pm$1.6 \\
\midrule
\multirow{1}{*}{Spatial} 
    & Stride/Cycle Length & 83.4\% & $\pm$13.3 \\
\bottomrule
\end{tabularx}
\end{table}

\paragraph{Temporal Parameters}
The Shapiro–Wilk normality test indicated that both cycle phase duration ($p > 0.095$) and swing phase duration ($p > 0.501$) were normally distributed, while stance phase duration was not ($p = 0.013$).
A paired t-test revealed a statistically significant difference between \ac{SFO} and treadmill for swing phase duration ($t(7) = 9.74$, $p < 0.001$), Cohen’s d~=~3.44 (large effect size), as well as for cycle phase duration ($t(7) = -18.7$, $p < 0.001$), Cohen’s d = 6.59 (large effect size), which was significantly shorter in SFO compared to treadmill.
Furthermore, a Wilcoxon signed-rank test on stance duration data showed no significant difference ($W = 8$, $p = 0.195$, with moderate effect $r = 0.50$).

A Pearson correlation analysis showed a strong and significant correlation for cycle duration ($r = 0.973$, $p < 0.001$), and a moderate but not statistically significant correlation for swing duration ($r = 0.673$, $p = 0.067$).
Additionally, a Spearman correlation analysis for stance duration, which was not normally distributed, did not reveal a significant correlation ($\rho = -0.167$, $p = 0.693$).

\paragraph{Spatial Parameters}
The Shapiro–Wilk test indicated that cycle length ($p = 0.259$) was normally distributed.
A paired t-test revealed a statistically significant difference in cycle length between the treadmill and orthosis conditions ($t(7) = -3.65$, $p = 0.008$), Cohen’s $d = 1.29$ (large effect), indicating that cycle length was significantly shorter when measured with the orthosis compared to the treadmill.
Pearson correlation analysis showed no significant correlation for cycle length ($r = 0.678$, $p = 0.065$).

\subsubsection{Pressure Data Measurement}
\begin{figure} [H]
\begin{adjustwidth}{-\extralength}{0cm}

    \centering
    \includegraphics[width=1\linewidth]{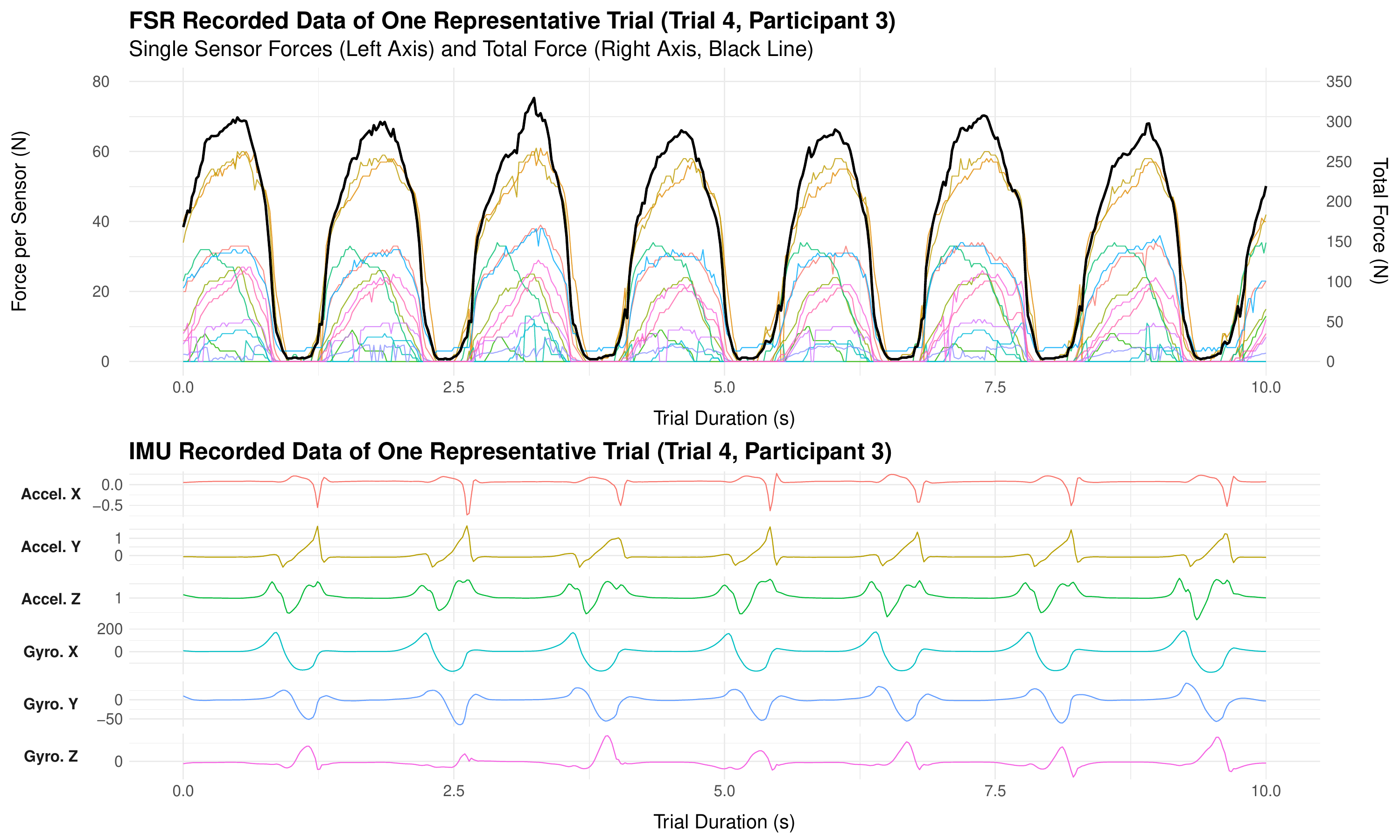}
    \end{adjustwidth}
    \caption{(\textbf{Top}): {Measured} 
 plantar pressure data of Trial 4 (P3), showing force values recorded by multiple FSR sensors over the duration of the task. Distinct gait phases can be identified from the characteristic patterns in the force signals. The left axis represents the force values per sensor, while the right axis indicates the total force (black line) across all included sensors.
(\textbf{Bottom}): Measured acceleration and gyroscope data from the IMU for each of the three axes, illustrating gait phase conformity with the recorded pressure data.}
    \label{Force-Data}
\end{figure}

The force data recorded with the \ac{SFO} enable the identification of gait cycles and their characteristic phases (initial contact, full contact, and toe-off), as illustrated in Figure~\ref{Force-Data}, demonstrating its suitability as a mobile gait analysis system.
However, a direct quantitative comparison with the instrumented treadmill is not feasible due to fundamental differences in measurement and data availability.
While the Zebris system measures contact forces between the treadmill surface~and the bottom of the orthosis, the \ac{SFO} records plantar forces directly between the foot and the insole. 
In addition, the Zebris software provides only aggregated outputs with averaged curves rather than continuous raw data, which prevents a synchronized step-wise comparison.

\begin{figure}[H]
    \begin{adjustwidth}{-\extralength}{0cm}
    \includegraphics[width=\linewidth]{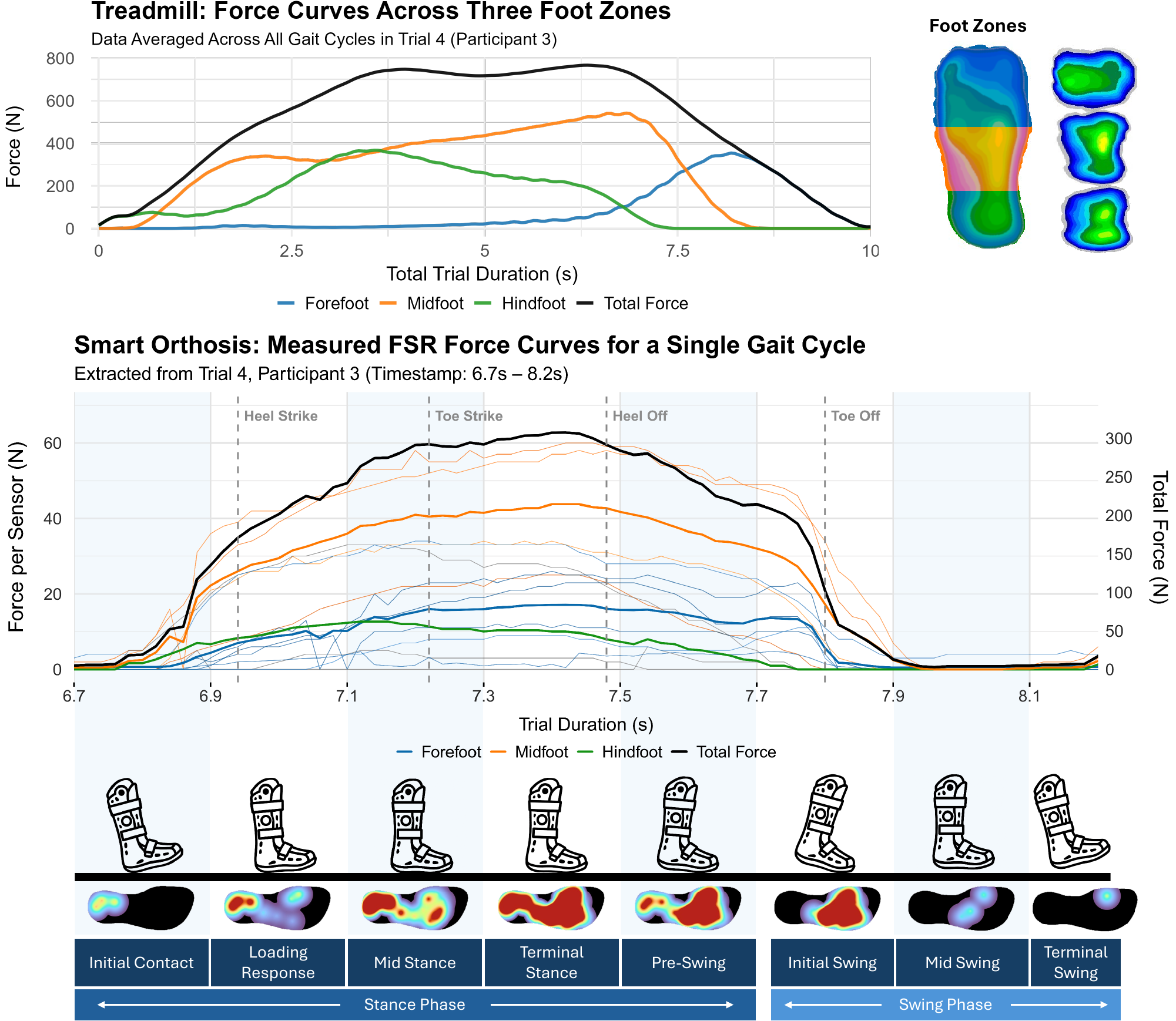}
    \end{adjustwidth}
    \caption{{The} 
 upper plot shows the aggregated plantar force curves of gait cycles recorded in Trial 4 of Participant 3 using the Zebris system. The force curves are color-coded according to three anatomical foot regions, as indicated by the footprint icon on the right. Additionally, a cumulative plantar pressure heatmap illustrates the spatial force distribution over the entire trial. The lower plot displays data from the same trial (Trial 4, Participant 3), focusing on a single gait cycle recorded by the orthosis. The three regional force curves represent the summed signals of the individual \acp{FSR} and are color-coded accordingly. Four gait events identified by the \ac{SFO} are marked in gray. Below the graph, individual heatmaps illustrate the pressure distribution for each gait event.}
    \label{GaitCycle}
\end{figure}

Nevertheless, to compare both systems a qualitative comparison of plantar pressure distributions and force curves was conducted using visualizations from both systems for a representative trial (see Figure~\ref{GaitCycle}). 
The Zebris data represent averaged force curves for the three foot zones summarized from multiple gait cycles, whereas the orthosis data focus on one representative gait cycle from the same trial, which is also displayed in Figure~\ref{Force-Data}.

Although absolute force values are not directly comparable due to differences in sensor configuration, both systems show similar temporal trends across the three anatomical foot zones and in the overall force–time diagram. 
Compared to the Zebris treadmill, which provides only aggregated contact data, the \ac{SFO} offers step-wise heatmap visualizations, enabling a detailed analysis of the internal plantar pressure distribution for each gait event. 

\subsection{Application Usability: Evaluation of Smartphone App Usability}
\label{4.2}
\begin{figure}[H]
\subfloat[\centering]{\includegraphics[width=0.4\linewidth]{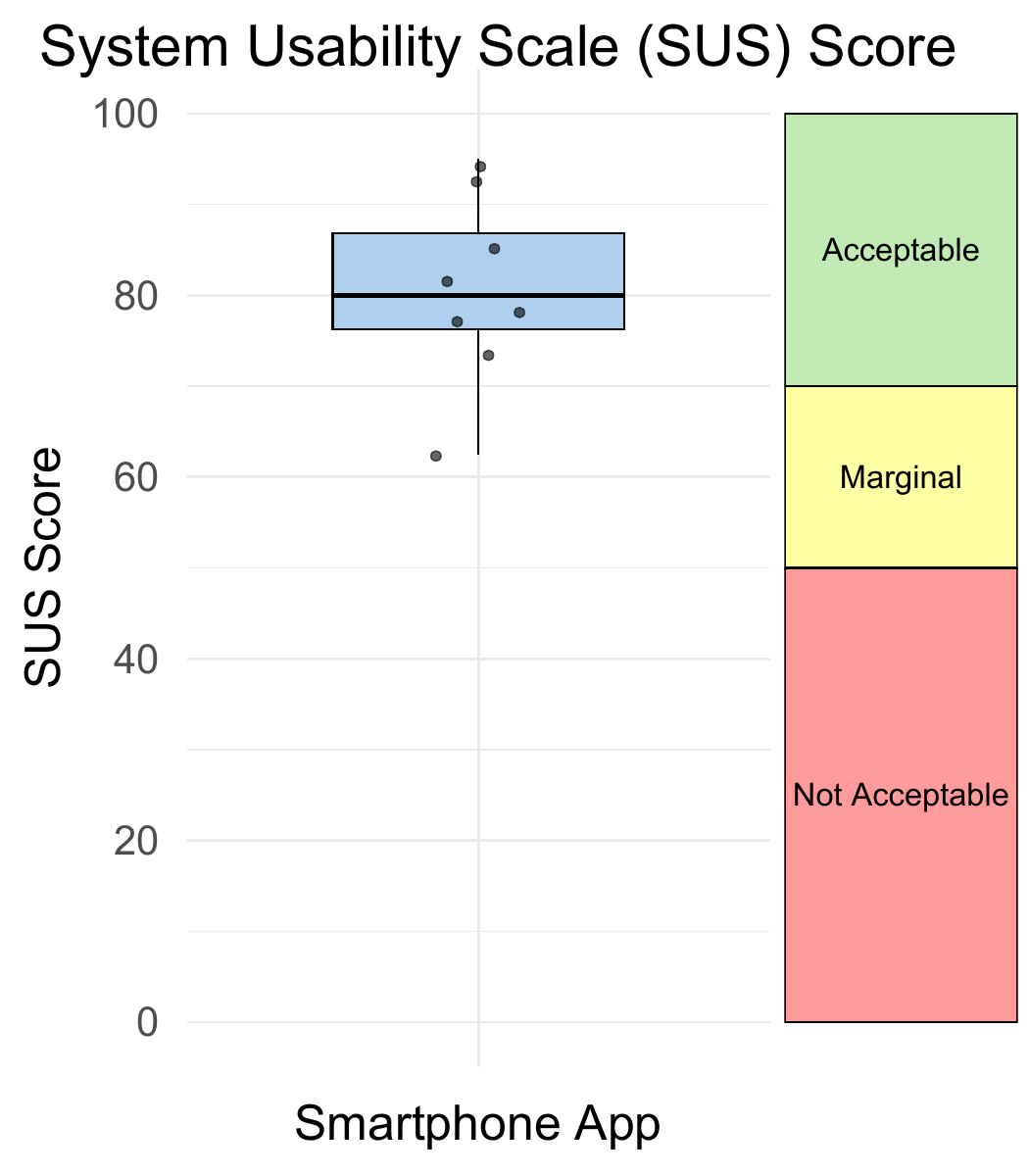}}
\hfill
\subfloat[\centering]{\includegraphics[width=0.58\linewidth]{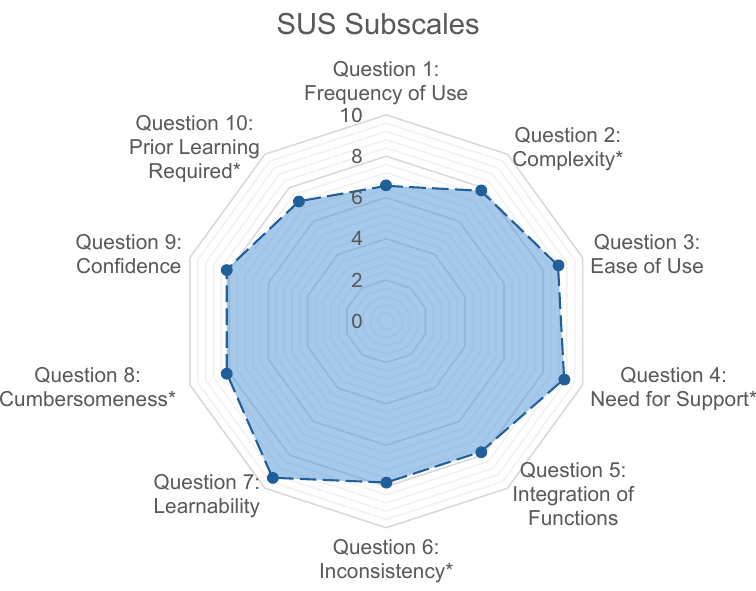}}\\
\caption{(\textbf{a}) Boxplot of overall SUS scores and the corresponding acceptability categories; (\textbf{b}) Radar plot visualizing the ten individual item scores, normalized on a 0–10 scale. Asterisks (*) indicate reverse-scored items.\label{figure9}}
\end{figure}

The mean \ac{SUS} score was 80.62 (\textit{SD} = 10.59), as shown in the boxplot diagram (\mbox{Figure~\ref{figure9}a}). 
According to the interpretation guidelines by Bangor et al.~\cite{SUS-Interpretation}, this score falls within the “Acceptable” range based on the Acceptability Ranges and is considered “Good” according to the Adjective Ratings, indicating a high level of usability.
Ratings for the ten individual \ac{SUS} items are visualized in the radar plot (Figure~\ref{figure9}b), with values ranging from 6.56 to 9.38 (normalized on a 0–10 scale).
Furthermore, the raw ratings of the subscales are summarized in Table~\ref{SUS-Score} using descriptive statistics.


%
%


\vspace{-6pt}

\begin{table}[H]
\caption{Descriptive statistics of the SUS items (M = Mean, SD = Standard Deviation).}
\label{SUS-Score}
\newcolumntype{C}{>{\centering\arraybackslash}X}
\begin{tabularx}{\textwidth}{ClCC} 
\toprule
\textbf{No.} & \textbf{Item} & \textbf{M} & \textbf{SD} \\
\midrule
1  & Frequency of Use              & 3.63 & 0.92 \\
2  & Complexity *                    & 1.88 & 0.83 \\
3  & Ease of Use                   & 4.50 & 0.53 \\
4  & Need for Support *              & 1.38 & 0.74 \\
5  & Integration of Functions      & 4.13 & 0.99 \\
6  & Inconsistency *                 & 1.88 & 0.64 \\
7  & Learnability                  & 4.75 & 0.46 \\
8  & Cumbersomeness *                & 1.75 & 1.04 \\
9  & Confidence                    & 4.25 & 0.71 \\
10 & Prior Learning Required *       & 2.13 & 1.36 \\
\bottomrule
\end{tabularx}

\footnotesize{\textit{Note.} Asterisks (*) indicate reverse-scored items.}
\end{table}

\subsection{Feedback Evaluation: User Perception of Haptic Feedback (Smart Orthosis and Walking Aid)}
\label{4.3}
Shapiro–Wilk tests were conducted for each condition on the questionnaire items \textit{Noticeability}, \textit{Non-Intrusiveness}, \textit{Regular Usage}, and \textit{Usefulness}, and revealed that the data were not normally distributed.
Therefore, an \ac{ART} \ac{RM} \ac{ANOVA} was conducted on the four measures, revealing two significant main effects.
A statistically significant main effect of \textit{Noticeability} was found for \textit{Device} with $F(1, 21) = 4.49$, $p = 0.046$, $\eta^2_p = 0.18$ (large effect size). 
Post hoc pairwise comparisons using Wilcoxon signed-rank tests did not reveal a significant difference between \textit{Orthosis} and \textit{Crutch} ($p = 0.065$).
However, the corresponding effect size ($r = 0.483$, moderate-to-large) suggests a potentially meaningful difference that may not have reached significance due to the limited sample size and resulting low statistical power.
Furthermore, a statistically significant main effect was found for \textit{Usefulness} on \textit{Feedback Type} with $F(1, 21) = 6.01$, $p = 0.023$, $\eta^2_p = 0.22$ (large effect size).
Post hoc Wilcoxon signed-rank tests revealed a significant difference between \textit{Continuous} and \textit{Pulsed Pattern} ($p = 0.008$).

\begin{figure} [H]
    \includegraphics[width=1\linewidth]{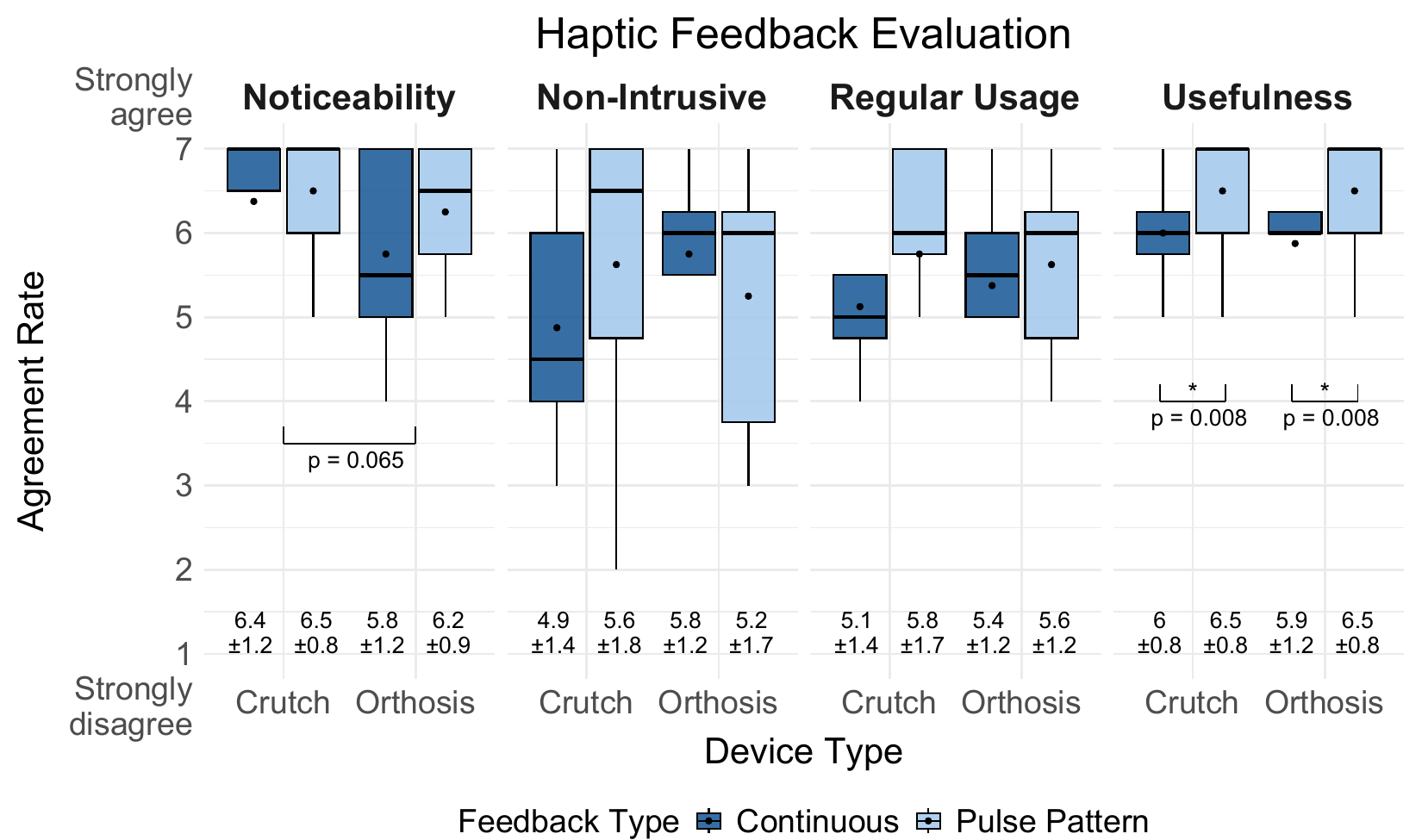}
    \caption{Boxplots show agreement ratings for four measures on haptic feedback via the foot orthosis and the forearm crutch, ranging from 1 (strongly disagree) to 7 (strongly agree). It displays the interquartile range (box), 1.5 × IQR range (whiskers), median (line), and mean (black dot). Mean and standard deviation (±) values are shown below each boxplot. Asterisks (*) indicate statistically significant results ($p < 0.05$).}
    \label{Feedback}
\end{figure}

Results of the \ac{ART} \ac{RM} \ac{ANOVA} and post hoc pairwise comparisons for all conditions are shown in Table~\ref{tab:anova_questionnaire}.
Additionally, the overall results are presented as a boxplot with descriptive statistics in Figure~\ref{Feedback}.

\vspace{-6pt}
\begin{table}[H]

\caption{Results of the ART RM ANOVA and post-hoc Wilcoxon signed-rank tests for the four questionnaire items.}
\label{tab:anova_questionnaire}
\begin{adjustwidth}{-\extralength}{0cm}
\centering
\newcolumntype{C}{>{\centering\arraybackslash}X}
\begin{tabularx}{\fulllength}{l l c c c c c c c c c} 
\toprule
 & & \multicolumn{5}{c}{\textbf{ART RM ANOVA}} & \multicolumn{4}{c}{\textbf{Pairwise Comparisons (Wilcoxon)}} \\
\cmidrule{3-7} \cmidrule(lr){8-11} 
\textbf{Measure} & \textbf{Factor} & \textbf{F} & \textbf{Df} & {\textbf{\textit{p}}} 
 & \textbf{Sig.} & \boldmath{$\eta^2_p$} & \textbf{Statistic} & \textbf{\textit{p}} & \boldmath{$r$} & \textbf{Magnitude} \\
\midrule
\multirow{3}{*}{\textit{Noticeability}} 
  & Device           & 4.49 & 1, 21 & 0.046 & * & 0.176 & 25.0 & 0.065 & 0.483 & moderate \\
  & Feedback         & 1.80 & 1, 21 & 0.194 &   & 0.079 & 22.0 & 0.187 & 0.310 & moderate \\
  & Device × Feedback & 1.60 & 1, 21 & 0.220 &   & 0.071 & — & — & — & — \\
\midrule
\multirow{3}{*}{\textit{Non-Intrusiveness}} 
  & Device           & 0.06 & 1, 21 & 0.815 &   & 0.003 & 27.5 & 0.653 & 0.106 & small \\
  & Feedback         & 0.50 & 1, 21 & 0.487 &   & 0.023 & 42.5 & 0.811 & 0.046 & small \\
  & Device × Feedback & 2.24 & 1, 21 & 0.150 &   & 0.096 & — & — & — & — \\
\midrule
\multirow{3}{*}{\textit{Regular Usage}} 
  & Device           & 0.00 & 1, 21 & 0.966 &   & 0.000 & 42.5 & 0.807 & 0.099 & small \\
  & Feedback         & 2.57 & 1, 21 & 0.124 &   & 0.109 & 50.0 & 0.129 & 0.326 & moderate \\
  & Device × Feedback & 0.77 & 1, 21 & 0.390 &   & 0.035 & — & — & — & — \\
\midrule
\multirow{3}{*}{\textit{Usefulness}} 
  & Device           & 0.00 & 1, 21 & 1.000 &   & 0.000 & 24.5 & 0.851 & 0.076 & small \\
  & Feedback         & 6.01 & 1, 21 & 0.023 & * & 0.220 & 36.0 & 0.008 & 0.703 & large \\
  & Device × Feedback & 0.00 & 1, 21 & 1.000 &   & 0.000 & — & — & — & — \\
\bottomrule
\end{tabularx}%
\end{adjustwidth}
\footnotesize{\textit{Note.} Asterisks (*) indicate statistically significant results ($p < 0.05$). Effect sizes are reported as partial $\eta^2$ for ART RM ANOVA and as rank-biserial correlation ($r$) for Wilcoxon signed-rank tests. 
Magnitude classification of $r$: small ($r<0.3$), moderate ($0.3\leq r<0.5$), large ($r\geq0.5$)~\cite{cohen2013statistical}.}

\end{table}

\subsubsection{Correlation of Ratings Across Devices}
To further investigate consistency in user perception across devices, a Spearman rank correlation was conducted between ratings for the same \textit{Feedback Type} on the \textit{Orthosis} and the \textit{Crutch}.
Results showed a significant positive correlation for \textit{Noticeability} in the \textit{Pulsed Pattern} feedback condition ($r_s = 0.86$, $p = 0.006$), suggesting a high level of consistency in perceived noticeability across both devices.
Furthermore, in the \textit{Continuous} feedback condition, a significant positive correlation was found for \textit{Regular Usage} ($r_s = 0.78$, \mbox{$p = 0.023$),} while the remaining conditions showed positive trends, but did not reach statistical significance: \textit{Noticeability} ($r_s = 0.62$, $p = 0.099$), \textit{Non-Intrusiveness} ($r_s = 0.56$, $p = 0.152$), and \textit{Usefulness} ($r_s = 0.58$, $p = 0.133$).

\subsubsection{Post-Experiment Questionnaire}
Participants rated the feedback types for both devices separately to identify preferences between the different modalities.
For the \textit{Orthosis}, the \textit{Pulsed Pattern} feedback was preferred by four participants, while the remaining four participants rated both feedback types as equally acceptable. 
None of the participants selected the \textit{Continuous} signal or no feedback as their preferred option.
For the \textit{Crutch}, six participants preferred the \textit{Pulsed Pattern} feedback.
Two participants indicated no preference and rated both options equally.
Similar to the orthosis results, no participant selected the \textit{Continuous} signal or no feedback.

\section{Discussion}
\label{sec:Discussion}
{
We developed a sensorized system for monitoring plantar pressure and gait activity within an orthopedic foot orthosis.
The system provides haptic feedback through either the orthosis or a corresponding forearm crutch and enables remote monitoring via a smartphone application.
In an experimental user study with eight participants, we evaluated the system’s feasibility and investigated the influence of device location and feedback modality on user perception of vibrotactile biofeedback.
The findings confirm the system’s technical feasibility for gait phase detection, the usability of the accompanying mobile application, and provide new insights into haptic feedback perception.
The following subsections discuss the results in detail, including implications, limitations, and directions for future~research.}

{\subsection{System Feasibility and Application Usability}}
{\subsubsection{System Feasibility: Accuracy of Plantar Pressure and Gait Parameters (SRQ1)}}
\ac{SRQ}1 focused on evaluating the accuracy of plantar pressure and gait measurements obtained from the developed \ac{SFO} compared to a clinical reference system.
Our findings showed that the recorded temporal parameters for the swing and cycle phases indicated significantly shorter durations when measured with the \ac{SFO} compared to the treadmill.
However, cycle duration showed a strong correlation between both systems, whereas swing duration did not reach statistical significance ($p = 0.067$).
Overall, swing duration showed the lowest accuracy among all temporal parameters, with only 60\% accuracy and notably higher standard deviations.
This may be explained by the observation that some \ac{FSR} sensors were constantly triggered and did not return to zero during offloading, which could have impaired the correct detection of events from stance to swing.
Although the stance phase showed no significant differences and achieved high accuracy of 96\%, the correlation between both systems was not significant.
These inconsistencies are likely caused by differences in event detection, as the \ac{SFO} relies on \ac{FSR}-based thresholds, whereas the treadmill reference system uses ground reaction forces.
Additionally, some triggering issues of certain \ac{FSR} sensors may have introduced inconsistent deviations in individual measurements, which could explain the reduced correlation, even though the average stance durations between both systems were similar.
As highlighted in previous research, lower walking speeds (0.8–1.0 m/s) compared to moderate walking speeds (1.2–1.4 m/s) are associated with higher error rates in temporal parameters (e.g., stance and swing time), as well as spatial parameters (e.g., stride length)~\cite{ganguly2023accuracy}.
As the walking speed in our experiment was relatively low (0.417 m/s) to ensure participant comfort while walking with the orthosis, this factor may have contributed to reduced accuracy in gait parameter~estimation.

The measured accuracy of the spatial parameter (cycle length) indicated that an accuracy above 83\% can be achieved.
However, the estimation of spatial gait parameters using a single smart insole systematically underestimated cycle length compared to treadmill-based reference measurements, and no significant correlation was found between both measurement systems.
One explanation for these systematically lower values is that the insole was worn unilaterally, while the other foot was not equipped with a sensorized insole to provide complementary data.
Furthermore, wrong triggering of the algorithm could have led to incorrect gait event detection.
By identifying this systematic offset, a post-processing correction factor could be applied to improve the dataset for further analyses.

The evaluation of pressure data showed that gait events can be reliably detected and continuous pressure distribution can be measured during both walking and standing.
Although the comparison with the treadmill was performed on a qualitative basis, the results showed that the three foot zones (forefoot, midfoot, and hindfoot) exhibited similar temporal patterns.
This highlights the added value of the \ac{SFO}, as it enables step-wise heatmap visualizations and detailed insights into the internal plantar pressure distribution without post-processing, which are not achievable with stationary treadmill systems.
In summary, our results confirm previous findings showing that one \ac{IMU} is sufficient for gait monitoring~\cite{MotivationIMU} and that the recommended configuration of 15 \acp{FSR}~\cite{15FSR-Sensors} is suitable for reliable pressure measurement.
However, further refinements are recommended to improve measurement accuracy and enhance data reliability.

{
Compared to related studies on smart insoles and gait monitoring systems, the accuracy levels achieved in this work are within a comparable range. 
For example, D’arco~et~al.~\cite{InsolesReview-ML} reported high performance in activity recognition using smart insoles, with accuracies ranging from 70\% to 99\%.
This demonstrates that comparable performance levels can be achieved even without additional post-processing.
Despite differences in sensor configuration (type, quantity, and placement), as well as varying application contexts, testing scenarios and data processing, the present study demonstrates that reliable gait analysis can also be achieved using a single insole.
With further iterative refinements of the sensing and calibration procedures, even higher accuracy levels may be achieved in future~implementations.}


{\subsubsection{Application Usability: Evaluation of Smartphone App Usability (SRQ2)}}
Regarding \ac{SRQ}2, which addressed the usability of the accompanying mobile application for visualizing gait and pressure data, our results showed an acceptable level of usability. 
The \ac{SUS} evaluation reached an average score of 80.62 (\textit{SD} = 10.59), which is comparable to the results reported by Resch et al.~\cite{Resch2024} for a smart insole app (type: basic), with a mean \ac{SUS} score of 77.5 (\textit{SD} = 16.74).
In comparison to other digital health applications, our findings also exceed the average usability benchmark identified by Hyzy et al.~\cite{Hyzy}, who conducted a meta-analysis and reported a mean \ac{SUS} score of 68.05 (\textit{SD} = 14.05) across various digital health apps.

Overall, the subscale ratings revealed that item 4 (\textit{need for support}) and item 7 (\textit{learnability}) received the highest scores.
This indicates that participants largely agreed with the statement ``\textit{most people would learn to use this app very quickly}'' (item 7) and disagreed with the statement ``\textit{I would need the support of a technical person to use the system}'' (item~4), which was reverse-scored.
In contrast, item 1 (\textit{frequency of use}) received the lowest average rating (\mbox{\textit{M} = 3.62}, \textit{SD} = 0.92 on a 5-point Likert scale), which shows that participants potentially were less willing to use the app frequently.
This might be due to the fact that some participants were only temporarily affected by foot-related conditions and therefore did not perceive frequent use of the app as necessary.
However, to address the potential limited interest in frequent use, future iterations of the app could incorporate targeted strategies to increase user engagement.
For example, personalized notifications and reminders could encourage consistent daily use, particularly for self-monitoring tasks such as the integrated pain-level diary.
Previous research suggested that gamification elements in health interventions (e.g., progress tracking or reward systems) can have positive effects on user experience~\cite{Gamification}.
Therefore, the integration of personalized measurable goals and feedback could support user motivation and foster continued use, especially in long-term rehabilitation contexts.\\

{\subsection{Feedback Evaluation: User Perception of Haptic Feedback (MRQ)}}
The \ac{MRQ} investigated how different haptic feedback types (\textit{Continuous} vs. \textit{Pulsed Pattern}) implemented in an orthosis and a walking aid affect user perception regarding \textit{Noticeability}, \textit{Non-Intrusiveness}, \textit{Regular Usage}, and \textit{Usefulness}.
Descriptive results showed that participants generally rated the \textit{Pulsed Pattern} feedback higher than the \textit{Continuous} feedback across all four items for both devices.
{A likely reason is that the rhythmic structure of the pulsed signal was perceived as more distinct and meaningful, making it easier to recognize and less monotonous compared to a continuous vibration.}
An exception was found for \textit{Non-Intrusiveness}, where the \textit{Pulsed Pattern} received lower ratings specifically in the orthosis condition.

Inferential analysis using an \ac{ART} \ac{RM} \ac{ANOVA} revealed a significant main effect of \textit{Device} on \textit{Noticeability}, as the crutch was rated significantly higher than the orthosis.
One possible explanation is that some participants had reduced sensation in the lower limbs, which made the haptic signal (especially \textit{Continuous} feedback) less perceptible when delivered via the foot.
Although previous work has suggested that foot-based feedback may be preferable to hand-based in some applications~\cite{FeedbackFootvsHand}, in this specific use case the hand appears to be more sensitive to the vibration signal using one \ac{ERM} actuator.
{This highlights that feedback effectiveness is not only dependent on signal type, but also on feedback location and tactile sensitivity at different body sites and devices.}

For \textit{Non-Intrusiveness}, the ratings varied more strongly across conditions.
In particular, the \textit{Continuous} feedback was rated lower than the \textit{Pulsed Pattern} when applied via the crutch,  which indicates that continuous vibrations may be perceived as more intrusive in this form factor.
The \textit{Regular Usage} showed similar ratings, however, without significance.
Ratings for perceived \textit{Usefulness} showed a significant effect of feedback type, indicating that the \textit{Pulsed Pattern} was perceived as more useful than the \textit{Continuous} feedback. 
{These findings highlight that temporally modulated vibration signals are more easily interpreted and less annoying than continuous signals in wearable biofeedback applications.}
This result aligns with the post-study feedback, in which most participants expressed a clear preference for the pulsed signal.
Although other trends were observed, they did not reach significance, which is probably due to the small size and limited statistical power of the~study.

These findings support previous research on the utility of \ac{ERM}-based vibration feedback~\cite{Vibration-ERM}. 
{However, our results extend this work by demonstrating that both pattern and device location influence the perception of vibrotactile feedback in orthopedic assistive devices.}
While a single \ac{ERM} actuator ensured sufficient noticeability in the crutch, we recommend using at least two actuators inside the orthosis to enhance perceptibility.
Overall, our results underscore the importance of tailoring haptic feedback implementation to the device form factor and user preferences.
{Therefore, designers should consider not only the feedback modality but also where and how it is delivered, and ensure that the signal is both interpretable and acceptable during everyday use.}
In line with prior work emphasizing the need to evaluate haptic feedback in assistive systems~\cite{technologies13080346}, our study contributes initial insights into user perception, acceptance, and effectiveness. \\

\subsection{Implications}
The presented system was designed as a general-purpose gait monitoring and feedback platform that can be applied both to individuals with foot conditions and for preventive or rehabilitative use in broader populations. 
While no clinical validation has yet been conducted, the system’s sensing and feedback capabilities hold potential for future application, particularly for monitoring plantar pressure distribution and detecting overload patterns during daily activities.

{The primary contribution of this work lies in advancing the understanding of how vibration-based feedback can be applied and perceived in foot-related assistive systems. 
While the developed hardware and software components served as necessary groundwork, they were mainly used for system verification. 
Building on this foundation, the central contribution is the investigation of real-time haptic overload notifications driven by the integration of plantar pressure and motion measurements.}
{The fused dataset from pressure and motion sensors enabled a closed-loop feedback mechanism that responds to gait events. 
Our results demonstrate that such multimodal data can be feasibly integrated to trigger vibration cues. 
Based on this, the evaluation of haptic feedback provided new insights into the perception and preferences related to vibration-based notifications in assistive systems. 
The findings show that pattern-based feedback is perceived as more effective for overload alerts, while perceptual differences between the orthosis and the walking aid highlight the need for differentiated actuator placement, intensity modulation, and an increased number of actuators to ensure sufficient noticeability. 
These insights contribute to a clearer understanding of multimodal feedback perception in foot augmentation systems and can inform design guidelines for future adaptive vibration strategies in assistive devices.}

{Overall, these findings extend existing knowledge on isolated prototype developments by combining system integration, technical validation, and user-centered evaluation.
These implications} can guide future research and the development of intelligent footwear, multimodal feedback solutions, and interactive mobile health applications, contributing to the fields of \ac{HCI} and biomedical engineering.

\subsection{Limitations}
The findings of our study are subject to several limitations. 
First, the evaluation primarily focused on technical development and validation under controlled laboratory conditions.
In particular, the validation was limited to the right foot, as the orthosis was consistently worn on the right side.
Moreover, the gait analysis was conducted using short treadmill walking sequences, which may not accurately reflect natural gait behavior during prolonged use or in real-world environments.
In addition, calibration of the \ac{FSR} sensors was performed based on manufacturer specifications without subject-specific mechanical calibration, potentially affecting the accuracy of absolute force estimation.
Potential variation in sensor placement and the lack of automated recalibration routines may further impact signal reproducibility and long-term stability in real-world scenarios.
Furthermore, comparability with treadmill-derived force values was limited, as only aggregated gait data were available from the treadmill, whereas the insole provided full raw data.
Similarly, the developed algorithm for gait event detection was not quantitatively compared with existing data-driven gait phase recognition methods, nor validated on publicly available datasets.
Finally, the small sample size limits the statistical power of the study and the generalizability of the observed effects, and no control group was included to provide a baseline comparison between participants with and without foot conditions.
{Nevertheless, the present work was designed as a proof-of-concept investigation with the primary focus on validating the technical feasibility and feedback functionality of the developed sensor system within the intended target group.
Consequently, the study focused on end users with foot conditions to ensure practical relevance.
Accordingly, the findings represent an initial technical evaluation conducted under controlled laboratory conditions and serve as a foundation for subsequent large-scale studies aimed at generalizing the results.}

\subsection{Future Work}
The modular system architecture, which integrates multimodal sensing and feedback, demonstrates how next-generation orthotic devices can move beyond passive monitoring toward interactive and adaptive support.
To address current limitations, future work should aim to evaluate the system under real-world conditions and in clinical contexts. 
{
In this context, future testing should include varied environmental conditions, such as uneven outdoor surfaces or stair walking, to evaluate the system’s robustness under realistic gait scenarios.
Such experiments will be essential for verifying performance outside controlled laboratory settings.
Another direction for future work is the extended testing of the smart insole with regard to material specifications and biocompatibility to ensure long-term durability and safe use in rehabilitation contexts.
Specifically, future research should include a larger and more diverse participant sample to allow for more comprehensive statistical analyses and improve the generalizability of the findings.
Expanding the sample size will also enable subgroup comparisons and allow the assessment of long-term performance and usability across different user populations.
Future studies should further include a control group of participants without foot conditions to establish a baseline for gait parameter comparison and system optimization.
Including such a reference group with healthy participants will allow for more comprehensive interpretation of gait-related outcomes and support system optimization.
Moreover, a case study involving patients with gait impairments could provide valuable insights into the system’s robustness, usability, and therapeutic potential.}

From a methodological perspective, future evaluations should also move beyond subjective usability scales and include behavioral and performance-based assessments, such as latency, energy consumption, correction rate, or changes in gait symmetry when interacting with the feedback system. 
Additionally, future work could benchmark the algorithm against state-of-the-art gait phase detection approaches and validate it on open-source datasets to enable objective performance comparison and generalization testing.

From a technical perspective, the integration of additional sensing modalities, such as temperature, \ac{GPS}, or physiological sensors, could further enhance system functionality and contextual awareness.
The modular architecture supports the flexible integration of such components for future developments.
Furthermore, the use of \ac{AI}-based approaches, including machine learning techniques such as support vector machines or artificial neural networks, could improve gait phase classification, anomaly detection, and event prediction.
These models may enhance the accuracy and robustness of gait parameter estimation, enable personalized and adaptive feedback mechanisms, and support new functionalities such as fall detection.

At the application level, the smartphone interface may be extended with features such as chatbots or gamified elements to support user-centered interpretation of health data and increase user engagement. 
Furthermore, the effectiveness of augmented feedback could be examined in user-centered training scenarios to evaluate its potential impact on motor learning and rehabilitation outcomes.
As a next step, we plan to test the orthosis for foot augmentation in virtual reality training environments, with a particular focus on investigating participant feedback interactions in immersive scenarios through visual feedback.
Additionally, a follow-up study is currently being planned to improve the smartphone application onboarding process using mixed reality, with the goal of enhancing users’ initial interaction, comprehension, and engagement.

\section{Conclusions}
\label{sec:Conclusion}
This work presents the development of an \ac{SFO} system combining pressure and motion sensing, haptic feedback via a connected walking aid, and an \ac{mHealth} application for real-time visualization. 
The proposed architecture enables the acquisition of gait and pressure data while providing immediate feedback to users.
Beyond its technical implementation, this work contributes to the growing field of assistive smart systems in healthcare by demonstrating how sensor-based feedback can be integrated into wearable and augmentative medical devices. 
Despite the limited testing in controlled environments, the system represents a step toward future patient-centered solutions that combine sensing, feedback, and data interpretation in daily-life contexts.
We encourage researchers and practitioners to further explore the integration of adaptive feedback and context-aware interaction in orthopedic and rehabilitation technologies to support more personalized and effective gait~interventions.

\vspace{6pt} 





\authorcontributions{Conceptualization, S.R.; methodology, S.R.; software, S.R., A.K., A.C., N.S., F.R., M.Z. and Y.S.; validation, S.R.; formal analysis, S.R.; investigation, S.R.; resources, S.R.; data curation, S.R.; writing---original draft preparation, S.R.; writing---review and editing, S.R. and D.S.-M.; visualization, S.R.; supervision, D.S.-M.; project administration, S.R.; funding acquisition, S.R. All authors have read and agreed to the published version of the manuscript.}

\funding{This research was funded by the Hessian Ministry of Science and Art-HMWK, Germany (FL1, Mittelbau), and supported by the RISE program of the German Academic Exchange Service~(DAAD).}

\institutionalreview{The study was conducted in accordance with the Declaration of Helsinki, and approved by the Institutional Ethics Committee of the German Society for Nursing Science (No. 23-027, 12 February 2024).}

\informedconsent{Informed consent was obtained from all subjects involved in the~study.}

\dataavailability{Data are unavailable due to privacy restrictions.}

\acknowledgments{We thank Diana Völz and Valentin Schwind for their support of our research.}

\conflictsofinterest{The authors declare no conflicts of interest. The funders had no role in the design of the study; in the collection, analyses, or interpretation of data; in the writing of the manuscript; or in the decision to publish the results.} 





\appendixtitles{yes} 
\appendixstart
\appendix
\section[\appendixname~\thesection]{Software Architecture and Data Processing---Smart Orthosis} \label{appendix:A}
{\subsection{Raw Data Acquisition}
Accurate and synchronized data acquisition is essential to ensure reliable downstream signal processing and gait detection. 
To ensure synchronization, data from the \acp{FSR} and the \ac{IMU} were recorded at a sampling rate of 100 Hz, consistent with values reported in related studies~\cite{SamplingRate, SamplingRate2, SamplingRate3, FSR-IMU-Position}.
Analog pressure signals from the \acp{FSR} were acquired via a 32-channel analog multiplexer. 
Each sensor was connected in a voltage divider circuit with a 10 k$\Omega$ resistor and the resulting signals were digitized using the microcontroller’s 12-bit ADC.
\ac{IMU} data was acquired via the \ac{I²C} interface and processed into triaxial acceleration (m/s\textsuperscript{2}) and angular velocity (rad/s) using the BMI270's internal processing unit.
To ensure temporal synchronization, each sample was assigned a unified timestamp.
All sensor readings were subsequently organized into structured data packets for further processing.}
{\subsection{Pre-Processing
}
Sensor data pre-processing was performed to improve signal quality and prepare the data for subsequent segmentation and analysis.
Due to manufacturing tolerances and environmental influences, \ac{FSR} signals required individual calibration to obtain accurate force measurements.
Each \ac{FSR} sensor provides a non-linear voltage output that decreases with increasing applied force.
The analog readings were first converted to voltage values, from which the resistance \( R_{\text{FSR}} \) was calculated using the voltage divider principle (\mbox{Equation~\eqref{eq:voltageDivider}}). 
Here, \( R_M \) is the reference resistor (10\,k$\Omega$), \( V^+ \) is the supply voltage, and \( V_{\text{out}} \) is the measured voltage across the FSR. 

\begin{equation}
V_{\text{out}} = \frac{R_M \cdot V^+}{R_M + R_{\text{FSR}}}
\label{eq:voltageDivider}
\end{equation}
Following the manufacturer's specification~\cite{interlink_fsr_guide}, the raw sensor output was used to estimate relative plantar forces without applying sensor-specific calibration.
To reduce high-frequency noise, a moving average filter was applied to the estimated force signals (Equation~\eqref{eq:averagefilter}).
\begin{equation}
\text{filteredForce}[i] = \frac{1}{N} \sum_{k=0}^{N-1} \text{fsrForce}[i-k]
\label{eq:averagefilter}
\end{equation}}

{The BMI270 includes internal configuration options for filtering and calibration, which were utilized to enhance signal quality.
The sensor was configured with a measurement range of $\pm4$ g for the accelerometer and $\pm2000^{\circ}/$s for the gyroscope to capture typical gait-related movements as well as rapid rotational changes.
Offset correction was automatically performed during initialization by fixing the \ac{IMU} in a stationary position to determine the static bias for each axis.
These offset values were removed from all subsequent accelerometer and gyroscope samples to ensure base-corrected measurements (Equations~\eqref{eq:IMU1}~and~\eqref{eq:IMU2}).
\begin{equation}
\text{accelCorrected}[i] = \text{accelRaw}[i] - \text{accelOffset}
\label{eq:IMU1}
\end{equation}
\begin{equation}
\text{gyroCorrected}[i] = \text{gyroRaw}[i] - \text{gyroOffset}
\label{eq:IMU2}
\end{equation}
Additionally, separate internal low-pass filters were applied to the accelerometer and gyroscope axes to remove high-frequency noise (Equation~\eqref{eq:lowpass}).
\begin{equation}
\text{filteredValue}[i] = \alpha \cdot \text{filteredValue}[i-1] + (1 - \alpha) \cdot \text{rawValue}[i]
\label{eq:lowpass}
\end{equation}
A filter coefficient of $\alpha = 0.9$ was chosen to effectively reduce high-frequency noise.
Pre-processing of the \ac{IMU} data through offset correction and low-pass filtering ensures that the acceleration and rotation data are precise and reliable.}

{\subsection{Signal
 Processing}
In the signal processing stage, pre-processed sensor data were further refined to reduce residual noise and enable robust detection of gait events. 
To improve signal quality, Kalman filters~\cite{kalman1960} were applied to both \ac{FSR} and \ac{IMU} signals. 
The Kalman filter was selected due to its proven suitability for real-time gait analysis in wearable systems, including \ac{IMU}-based orientation estimation, walking distance calculation, and general gait assessment~\mbox{\cite{Review-GaitAnalysis, KalmanFilter-GaitAssessment, ExtendedKalmanFilter}}.
For the \ac{FSR} signals, individual Kalman filters were assigned to each of the sensor channels, with parameters set to $Q = 1.0$, $R = 1.0$, and $P = 0.01$. 
The filter parameters were adjusted to maintain responsiveness while reducing high-frequency fluctuations. 
For the \ac{IMU}, each axis of the accelerometer and gyroscope data was filtered individually, with configuration parameters adjusted to the sensor's noise characteristics.
Filter parameters were empirically set to $Q = 0.1$, $R = 0.1$, and $P = 0.3$ for accelerometer data, and to $Q = 1.0$, $R = 1.0$, and $P = 0.01$ for gyroscope data to balance signal smoothness and responsiveness.}

{The filtered \ac{FSR} signals served as the basis for gait event detection. 
A custom rule-based algorithm was implemented to identify key gait phases using threshold values defined as relative percentages of the user's body weight.
The algorithm evaluated pressure variations within specific sensor zones corresponding to the forefoot and hindfoot regions to detect events such as heel strike, toe strike, heel off, and toe off.
This individualized approach enabled the detection of characteristic loading patterns and supported the identification of critical gait events and potential overload conditions.
Thresholds were empirically determined through pre-testing across multiple gait cycles and optimized to minimize detection errors.
A finite state machine was implemented to manage gait phase transitions based on these events. 
This approach ensured a consistent event sequence and enabled accurate temporal segmentation.
A baseline correction algorithm was used to counteract the hysteresis effects of the \ac{FSR} sensors and to ensure zero-base alignment after the load is removed. 
This adjustment was essential for maintaining continuous and reliable gait event detection across steps.
The detected gait events, including their timestamps, were subsequently used to compute spatiotemporal gait parameters such as step duration, stance/swing phase duration, and cadence.}

{\subsection{Sensor
 Fusion}
Filtered \ac{FSR} and \ac{IMU} data were fused to extract both temporal and spatial characteristics of the gait cycle.
While \ac{FSR} data provide reliable information on foot-ground contact timing, the \ac{IMU} captures the foot’s three-dimensional orientation and motion dynamics. 
A Mahony filter~\cite{Mahony-Filter} was employed to estimate foot orientation in terms of roll, pitch, and yaw. 
The Mahony filter was implemented via {the} 
 \texttt{fusion.MahonyUpdate()} function from the SensorFusion library~\cite{library}.
This approach combines gyroscope data for short-term tracking with accelerometer data for long-term correction, effectively compensating for drift and high-frequency noise.
To improve the accuracy of position estimation, zero-velocity updates~\cite{ZeroVelocityDetection} were applied during the stance phase, identified from \ac{FSR} signals. 
These updates reset the velocity to zero between the heel contact and the heel off, thereby limiting integration drift and reducing cumulative error in position calculation.}

{\subsection{Extraction
 of Gait Parameters}
Based on the fused sensor data, key gait parameters were extracted. 
The segmentation of the gait cycle follows the definition of Perry~\cite{GaitPhasesDefinition-Perry}, which distinguishes two main phases: the stance phase (60\% of the gait cycle) and the swing phase (40\%), as well as eight sub-phases (including foot strike, foot flat, heel off, and foot off).
These gait parameters can be classified into spatial, temporal, spatio-temporal, and kinetic categories~\cite{GaitPhases}.
Stride length was calculated by integrating acceleration data, corrected for foot orientation and stabilized using zero-velocity updates during stance phases.
Gait speed was subsequently derived as the ratio of stride (cycle) length to gait cycle duration.
As the measurement setup included only a single foot-mounted \ac{IMU}, step length and step width could not be directly calculated.
However, when the instrumented insoles are worn on both feet, these parameters can be extracted.
In addition to standard parameters, a stability index was introduced to quantify gait stability. 
This metric integrates two normalized components:
\begin{itemize}
    \item \textls[-25]{Pressure variability: Temporal fluctuations in total plantar pressure across all \ac{FSR}~channels.}
    \item Anterior-posterior pressure ratio: Relative distribution of pressure between forefoot and rearfoot regions.
\end{itemize}
The final stability index was computed as a weighted sum of both components and ranges from 0 (low stability) to 1 (high stability). 
Higher values indicate low pressure variability and a balanced anterior-posterior distribution, whereas lower values suggest unsteady gait characteristics.
By combining \ac{FSR} and \ac{IMU} data, the system enables accurate, real-time estimation of gait events and parameters. 
The integration of Mahony filtering and zero-velocity detection effectively mitigates drift and noise, while the proposed stability index provides an interpretable measure for detecting variations in gait control.}

{The extracted gait parameters serve as the basis for subsequent data analysis and interpretation. 
Quantitative measured variables such as stride length, cadence, stance time and gait speed can be statistically evaluated.
Furthermore, the parameters enable the classification of gait patterns and the detection of anomalies, which may be indicative of pathological gait or early-stage motor impairments. }

\begin{adjustwidth}{-\extralength}{0cm}

\reftitle{References}

\PublishersNote{}
\end{adjustwidth}
\end{document}